\documentclass[twocolumn]{aastex7}

\newcommand\gaia{{\em Gaia\ }}
\newcommand\rguide{$R_{Guide}$~}
\newcommand\rgc{$R_{GC}~$}
\newcommand\occam{OCCAM~}
\newcommand\natp{\citet{myers_2022}}

\newcommand\sm{M{\'e}sz{\'a}ros~}

\usepackage{amsmath}
\usepackage{moresize}
\usepackage{soul}
\usepackage{gensymb}
\received{\today}

\submitjournal{AJ}

\shorttitle{OCCAM VIII.: Abundance Gradients from SDSS-V/MWM DR19}
\shortauthors{J.~M.~Otto, et al.}

\begin{document}

\title{The Open Cluster Chemical Abundances and Mapping Survey: VIII. \\ Galactic Chemical Gradient and Azimuthal Analysis from SDSS/MWM DR19}

\correspondingauthor{Jonah M.~Otto}

\author[0000-0003-2602-4302]{Jonah M.~Otto}
\affiliation{Department of Physics and Astronomy, Texas Christian University, TCU Box 298840 
Fort Worth, TX 76129, USA }
\email[show]{j.otto@tcu.edu}

\author[0000-0002-0740-8346]{Peter M.~Frinchaboy}
\affiliation{Department of Physics and Astronomy, Texas Christian University, TCU Box 298840 
Fort Worth, TX 76129, USA }
\affiliation{Maunakea Spectroscopic Explorer, Canada-France-Hawaii-Telescope, Kamuela, HI 96743, USA}
\email{p.frinchaboy@tcu.edu}

\author[0000-0001-9738-4829]{Natalie R. Myers}
\affiliation{Department of Physics and Astronomy, Texas Christian University, TCU Box 298840 
Fort Worth, TX 76129, USA }
\email{n.myers@tcu.edu}

\author[0000-0002-6534-8783]{James W.~Johnson}
\affiliation{Carnegie Science Observatories, 813 Santa Barbara Street, Pasadena, CA 91101, USA}
\email{jjohnson10@carnegiescience.edu}

\author[0009-0000-4049-5851]{John Donor}
\affiliation{Department of Physics and Astronomy, Texas Christian University, TCU Box 298840 
    Fort Worth, TX 76129, USA }
\email{j.donor@tcu.edu}

\author[0009-0008-6057-5339]{Ahabar Hossain}
\affiliation{Department of Physics and Astronomy, Texas Christian University, TCU Box 298840 
Fort Worth, TX 76129, USA }
\email{AHABAR.HOSSAIN@tcu.edu}

\author[0000-0001-8237-5209]{Szabolcs~M{\'e}sz{\'a}ros}
\affiliation{ELTE E\"otv\"os Lor\'and University, Gothard Astrophysical Observatory, 9700 Szombathely, Szent Imre H. st. 112, Hungary}
\affiliation{MTA-ELTE Lend{\"u}let "Momentum" Milky Way Research Group, Hungary}
\email{meszi@gothard.hu}

\author[]{Hailey Wallace}
\affiliation{Department of Physics and Astronomy, Michigan State University, East Lansing, MI 48824, USA}
\affiliation{Department of Physics and Astronomy, Texas Christian University, TCU Box 298840 
Fort Worth, TX 76129, USA }
\email{walla462@msu.edu}

\author[0000-0001-6476-0576]{Katia Cunha}
\affiliation{Observat{\'o}rio Nacional, Rua General Jos{\'e} Cristino, 77, Rio de Janeiro, RJ 20921-400, Brazil}
\affiliation{Steward Observatory, University of Arizona, 933 North Cherry Avenue, Tucson, AZ 85721-0065, USA}
\email{kcunha@arizona.edu}

\author[0000-0002-7707-1996]{Binod Bhattarai}
 \affiliation{Department of Physics, University of California, Merced, 5200 North Lake Road, Merced, CA 95343, USA}
\email{bbhattarai@ucmerced.edu}

\author[0000-0001-6761-9359]{Gail Zasowski}
 \affiliation{Department of Physics \& Astronomy, University of Utah, 115 S. 1400 E., Salt Lake City, UT 84112, USA}
\email{u0948422@gcloud.utah.edu}

\author[0000-0001-6761-9359]{Sarah R. Loebman}
 \affiliation{Department of Physics, University of California, Merced, 5200 North Lake Road, Merced, CA 95343, USA}
\email{sloebman@gmail.com}

\author[0009-0008-0081-764X]{Alessa I.~Wiggins}
\affiliation{Department of Physics and Astronomy, Texas Christian University, TCU Box 298840 
    Fort Worth, TX 76129, USA }
\email{A.IBRAHIM@tcu.edu}

\author[0000-0003-0872-7098]{Adrian M.~Price-Whelan}
\affiliation{Center for Computational Astrophysics, Flatiron Institute, 162 Fifth Avenue, New York, NY 10010, USA}
\email{aprice-whelan@flatironinstitute.org}

\author[0000-0003-4019-5167]{Taylor Spoo}
\affiliation{Department of Physics and Astronomy, Texas Christian University, TCU Box 298840 
    Fort Worth, TX 76129, USA }
\email{T.SPOO@tcu.edu}

\author[0000-0002-7883-5425]{Diogo Souto}
\affiliation{Departamento de F{\'i}sica, Universidade Federal de Sergipe, Av.~Marcelo Deda Chagas, S/N Cep 49.107-230, S{\~a}o Crist{\'o}v{\~a}o, SE,
Brazil}
\email{diogosouto@academico.ufs.br}



\author[0000-0002-3601-133X]{Dmitry Bizyaev}
\affiliation{Apache Point Observatory and New Mexico State University, P.O. Box 59, Sunspot, NM, 88349-0059, USA}
\affiliation{Sternberg Astronomical Institute, Moscow State University, Moscow, 119234, Russia}
\email{dmbiz@apo.nmsu.edu}

\author[]{Kaike Pan}
\affiliation{Apache Point Observatory and New Mexico State University, P.O. Box 59, Sunspot, NM, 88349-0059, USA}
\email{kpan@apo.nmsu.edu}

\author[orcid=0000-0002-6561-9002,gname=Andrew,sname=Saydjari]{Andrew~K.~Saydjari}
\altaffiliation{Hubble Fellow}
\affiliation{Department of Astrophysical Sciences, Princeton University, Princeton, NJ 08544 USA}
\email{aksaydjari@gmail.com}

\begin{abstract}
The Open Cluster Chemical Abundances and Mapping (OCCAM) survey seeks to curate a large, comprehensive, uniform dataset of open clusters and member stars to constrain key Galactic parameters. This eighth entry from the OCCAM survey, based on the newly released SDSS-V/MWM Data Release 19 (DR19), has established a sample of 164 high quality open clusters that are used to constrain the radial and azimuthal gradients of the Milky Way.
The DR19 cluster sample [Fe/H] abundances are roughly consistent with measurements from other large-scale spectroscopic surveys. However, the gradients we calculate deviate considerably for some elements. We find an overall linear Galactic radial [Fe/H] gradient of $-0.075 \pm 0.006 \text{ dex kpc}^{-1}$ using the cluster's current Galactocentric Radius (\rgc) and a gradient of $-0.068 \pm 0.005 \text{ dex kpc}^{-1}$ with respect to the cluster's guiding center radius. 
We do not find strong evidence for significant evolution of the differential element gradients ([X/Fe]) investigated here (O, Mg, Si, S, Ca, Ti, Cr, Mn, Fe, Co, Ni, Na, Al, K, Ce, Nd). 
For the first time using the OCCAM sample we have sufficient numbers of clusters to investigate Galactic azimuthal variations. In this work, we do find evidence of azimuthal variations in the measured radial abundance gradient in the Galactic disk using our open cluster sample.

\end{abstract}



\section{Introduction} \label{sec:intro}


Open clusters are a fundamental building block of galactic disks, making them an excellent tool for Galactic archaeology. 
In the last two decades, large spectroscopic surveys such as the Apache Point Observatory Galactic Evolution Experiment \citep[APOGEE,][]{apogee}, GALactic Archaeology with Hermes \citep[GALAH,][]{galah}, \gaia-ESO \citep[][]{gaia_eso}, and Large sky Area Multi-Object fiber Spectroscopic Telescope \citep[LAMOST,][]{lamost}, 
have made it possible to build upon the work of \citet{janes_79} to further explore and constrain the Galactic chemical gradient for a variety of element species, including $\alpha$, iron-peak, odd-z and neutron capture.  
Numerous studies \citep[e.g.,][]{yang2025, occasoV, magrini_23, gaia_grads, myers_2022, spina_21, occam_p4, magrini_2017, occam_katia, frinchaboy_13} have utilized these surveys and proven open clusters to be crucial barometers to study stellar evolution, calibrate key astronomical parameters, and trace Galactic chemical evolution. The majority of these studies have fit the radial metallicity gradient using a two-component fit with a free parameter transition point \citep{yang2025,occasoV,magrini_23,myers_2022,occam_p4}, but whether a bilinear fit out performs a single linear fit is still up for debate. 


The observational effort to define the Galactic metallicity gradient provides key constraints to the different chemical evolution models that have been proposed. 
Among them is the two-infall model, where one infalling gas cloud creates a halo and thick disk and a second, prolonged infall, forms the thin disk \citep[][]{1997ApJ...477..765C}.
This model tracks metallicity gradients and abundance-age relations along with radial variations \citep{2001ApJ...554.1044C}. In addition, \citet{1999MNRAS.307..857B} proposed an inside out disk formation model where they concluded that star formation occurs faster in the inner disk, indicating steeper metallicity gradients earlier in the Galaxy's history. 
Another model, utilizing radial migration and chemodynamics, combined cosmological simulations with chemical evolution prescriptions \citep{2014A&A...563A..16N} and predicted the flattening of radial gradients, a lack of a correlation between age and metallicity, and abundance trends across the galactic disk. 


Open clusters are ideal tracers of chemical abundance patterns for a variety of reasons.
They are coeval \citep{1995ARA&A..33..381F}; the stellar population within open clusters form nearly simultaneously from the same molecular cloud, allowing for a more precise determination of the ages of stars. They are also largely chemically homogeneous, with a typical scatter of $\sim$0.02 dex or less in most elements as seen in observational studies \citep[e.g.,][]{sinha_24,2016ApJ...817...49B}, as well as in simulated FIRE galaxies \citep[e.g.,][]{bhattarai_24}.

The last open cluster membership entry in the Open Cluster Chemical Abundance and Mapping (OCCAM) survey, \citet{myers_2022}, based on SDSS-IV/APOGEE DR17 \citep{dr17}, included 153 clusters, 94 of which were deemed high quality. As the number of clusters available for analysis increases, we can begin to investigate the nature of these azimuthal variations across the Galactic disk. 
Recent studies, \citep[e.g.,][]{hackshaw_2024,hawkins_23,poggio_22} have investigated these azimuthal variations using field giant stars. \citet{hawkins_23} and \citet{poggio_22} found evidence that the variations are correlated with the spiral arm structure. However, \citet{hackshaw_2024} found that the connection between azimuthal variations in the metallicity gradient and spiral arm structure to be inconclusive. 
Follow up studies leveraging the ever-growing open cluster sample will augment the field star work and help to clarify the ambiguities still present. 



In this paper, we present a larger, improved OCCAM sample of 1083 member stars in 164 open clusters, a 74\% increase in the number of `` high quality'' clusters over our previous OCCAM work \citep{myers_2022}.
In \S \ref{sec:data} we outline the data utilized for this study. In \S \ref{sec:methods} we detail the methodology used to complete our work. In \S \ref{dr19} we describe in detail the OCCAM sample and the available Value Added Catalog (VAC). Results are presented in \S \ref{sec:results}, discussed in \S \ref{sec:discussion}, with conclusions in \S \ref{sec:conclusions}. 



\section{Data} \label{sec:data}
This work leverages the stellar parameters and abundances from the Milky Way Mapper (MWM) (J. A. Johnson et al., {\em in prep}) survey as part of the 5th iteration of the Sloan Digital Sky Survey (SDSS-V; Kollmeier et al., {\em submitted}) as well as positional and kinematic data from the ESA \gaia mission \citep{gaia_tot}. By combining data from only two sources we create a uniform sample that minimizes any systematic offsets which arise when data is pulled from multiple sources. 

\subsection{SDSS-V/MWM DR19}

The 19th data release from SDSS-V (DR19; Aghakhanloo et al., {\em submitted}) includes chemical abundances and radial velocities from MWM for $\sim 1$ million stars. 
New high-resolution ($R\sim$ 22,500), near-infrared spectra have been taken with the APOGEE spectrographs \citep{wilson_2019} which have been added to the previous SDSS/APOGEE data from DR17 \citep{dr17} that has been re-reduced in a consistent manner using the APOGEE data reduction pipeline \citep[][D. L. Nidever et al., {\em in prep}]{nidever_2015}. These data were taken with the Sloan Foundation telescope at the Apache Point Observatory in New Mexico \citep{sloan_telescope} and the Du Pont telescope at the Las Campanas Observatory in Chile \citep{du_pont}, which provide coverage of the Northern and Southern Hemispheres, respectively. Detailed targeting information for the MWM survey can be found in \citet{dr18}, while targeting details for the APOGEE survey \citep[]{apogee} are found in \citet{frinchaboy_13}, \citet{zasowski13}, \citet{zasowski17}, \citet{ap2n_target} and \citet{ap2s_target}. Individual element abundances used in this study were derived using the APOGEE Stellar Parameters and Abundances Pipeline (ASPCAP) \citep[][]{aspcap}, which was run as part of the {\tt astra} (Casey et al., {\em in prep}) framework. 


\section{Methods} \label{sec:methods}

Individual star membership probabilities for the open clusters presented in this analysis were determined by first selecting SDSS-V/MWM Data Release 19 (DR19; Aghakhanloo et al. 2025, {\em in prep}) stars within $3\times R_{50}$\footnote{The radius containing half of the member stars as determined by \citet{cg20}} for each cluster. 
To ensure reliable membership, we select those stars that have a $>70\%$ membership probability in the open cluster catalog from \citet{cg20}, which uses the positions, proper motions, and parallaxes from \gaia DR2 to constrain cluster membership. 
By combining these filtered cluster members with the available and reliable RVs and metallicities from MWM, we are able to further constrain cluster membership and create a purer sample of member stars. 
To compute the RV and metallicity memberships for each cluster, we employ a method similar to that used in \citet{donor_18, occam_p4} and \citet{myers_2022}, wherein we apply a Gaussian kernel smoothing routine in first radial velocity (RV) space, then [Fe/H] space, for the stars which passed the earlier cuts (i.e., proper motion then RV). 
By fitting a Gaussian to this distribution and normalizing it, we are able to compute the membership probabilities for each star in each parameter space. 
For consistency with previous OCCAM papers, we present open clusters and members from \citet{cg20}, combined with the MWM/APOGEE radial velocities and ASPCAP \citep[]{aspcap} chemical abundances in the main body of this paper. 
Appendix \ref{ehmems} shows the results when the \citet{HUNT_2023} open cluster catalog is used as the starting point for the above procedure instead of the \citet{cg20} catalog\footnote{We prioritized the \citet{cg20} catalog over the \citet{HUNT_2023} because although \citet{HUNT_2023} is a newer, larger catalog, there are identified issues with the \citet{HUNT_2023} ages which are significantly systemically too young for the old open clusters.  This difference in age is due to the isochrone fitting method confusing blue stragglers for the main sequence turn off stars, resulting in incorrectly younger ages being measured.    
  Additionally, \citet{cg20} and \citet{HUNT_2023} provide mostly similar membership at the magnitudes we have spectra for, though \citet{HUNT_2023} does add additional membership for fainter stars that will be important for future deeper spectroscopic analysis/surveys.}.  

\begin{figure*}[t!]
 	\begin{center}
         \epsscale{1.05}
     \plotone{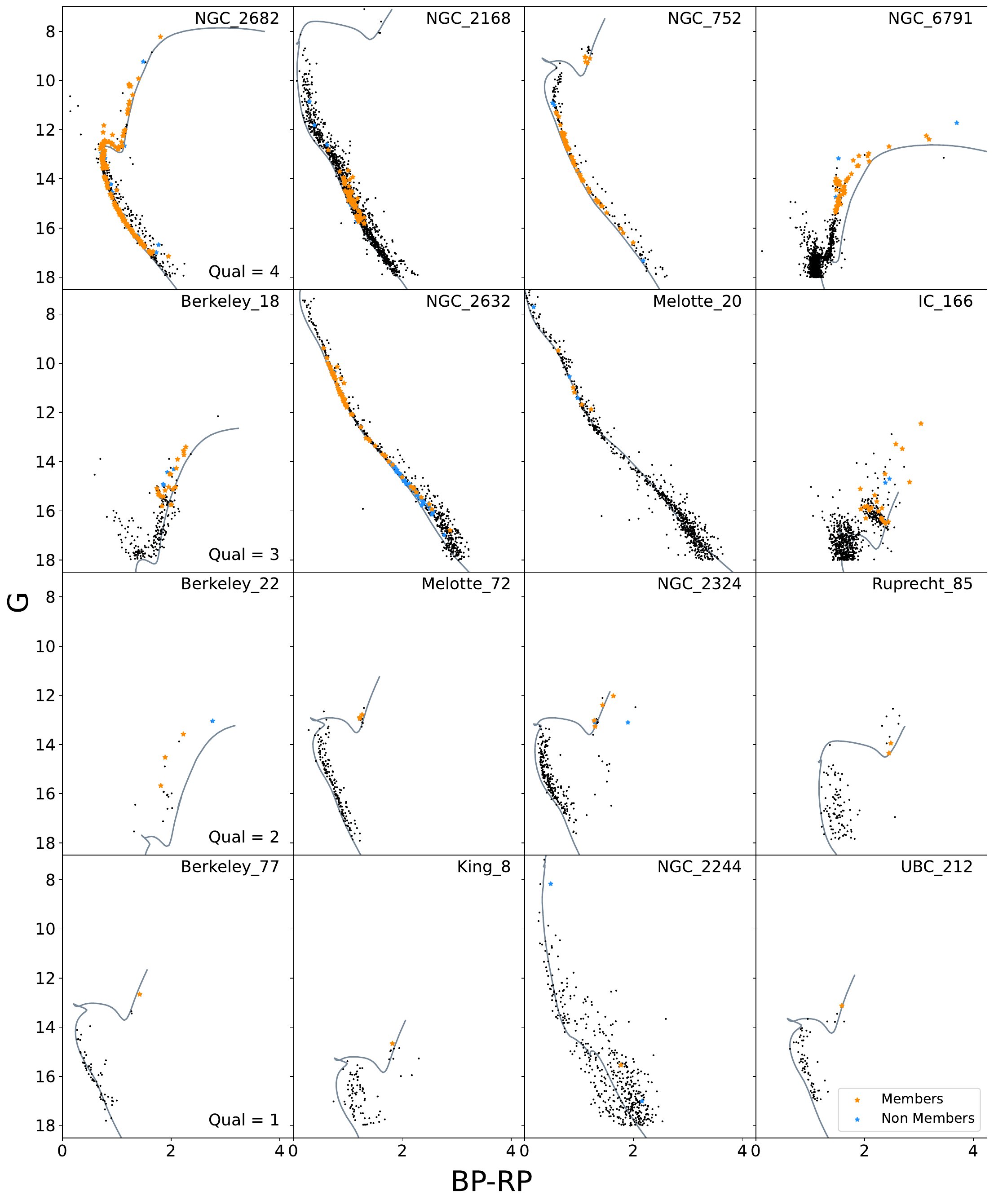} 
 	\end{center}
 	\caption{ \small A selection of example ESA \gaia color-magnitude diagrams ($BP-RP$, $G$) from each of the 4 quality categories. APOGEE/MWM calibration clusters ({\tt Qual = 4}) are shown in the first row, high quality clusters ({\tt Qual = 3)} with more than 5 stars in the second, high quality clusters ({\tt Qual = 2}) with 2-4 stars in the third, and good clusters ({\tt Qual = 1}) with only 1 star in the bottom row.  \citet{cg20} ESA \gaia-identified, proper motion-selected member stars are shown as black points, OCCAM pipeline-identified MWM/APOGEE members from DR19, adding RV and [Fe/H] selection, are shown as orange stars. The blue stars are proper motion-selected member stars that the OCCAM pipeline has rejected as RV and [Fe/H] members. A PARSEC isochrone generated with the mean cluster [Fe/H] from OCCAM, and the distance, reddening, and age from \citet{cg20} is plotted in grey. ({\it Note: No effort has been made to adjust/refit the isochrone fit in this work}). } 
 	\label{fig:cmds}
 \end{figure*}

As in previous OCCAM papers, a visual quality check was employed. Both Kiel diagrams ($T_{eff}$ vs log(g)) and color-magnitude diagrams (CMDs), with PARSEC isochrones \citep[]{bressan12-parsec} over-plotted, were used to determine the cluster quality. In this work, we categorize each cluster into one of five categories: calibration clusters ({\tt Qual = 4}), high quality clusters with 5$+$ members ({\tt Qual = 3}), high quality clusters with 2--4 members ({\tt Qual = 2}), good clusters with only one star ({\tt Qual = 1}), and rejected clusters. 
The clusters designated as calibration clusters represent the well-studied set of clusters first presented in \citet{donor_18}, along with additional open clusters specifically targeted by SDSS. 
All clusters with calibration, high quality, or good designations have isochrone fits that match well with the associated CMD, and are used in the gradient analysis.  
Rejected clusters could either be groups of stars that we do not believe are real clusters based on their isochrone and CMD combination, or are clusters that only have a single star with SDSS/MWM data that we do not believe to be a member of the cluster. We show example CMDs for a selection of clusters that fall into each of the four accepted categories in Figure \ref{fig:cmds}. 

The guiding center radius ($R_{Guide}$) is used as the primary radius, rather than $R_{GC}$, for this work. For any general orbit, \rguide is the radius of a perfectly circular orbit that has the same angular momentum as the eccentric orbit. \rguide was computed for every cluster in the sample using the circular velocity rotation curve from the 2022 Milky Way potential model in the {\tt gala} software package \citep[][]{gala}, a Galactic Dynamics code. Gradients with respect to both \rguide and Galactocentric radius ($R_{GC}$) were computed in this work, but only gradients with respect to \rguide are shown in the vast majority of the figures contained in this work. Several studies \citep[e.g.,][]{Netopil21,spina_21,Zhang21} have shown that using \rguide instead of the clusters present-day Galactocentric radius can help correct orbital blurring effects seen in chemical abundance gradients. In addition to $R_{Guide}$, the {\tt gala} software was used to calculate several different orbital parameters (current/max height above the plane, azimuth angle, eccentricity, average radial/$Z$ period) for each of the 164 clusters in our sample.

With the final sample of clusters, we preform both a linear and a bilinear fit of the metallicity ([Fe/H] gradient using the {\tt emcee} \citep[][]{emcee}, a Markov Chain Monte Carlo (MCMC) method software package. 
The bilinear fit divides the sample into two regions and fits lines to the data points within each region with the restriction that the two lines must intersect at a "knee", which is itself a free parameter. 
The resulting function can be expressed as:

\begin{equation}
y= \begin{cases}m_1 \cdot x+b_1 & x \leq k \\ m_2 \cdot(x-k)+\left(m_1 \cdot k+b_1\right) & x>k\end{cases}
\end{equation}

\noindent Values for the parameters $m_1$, $b_1$, $m_2$, and $k$ were estimated using maximum likelihood estimation, and uncertainties in each of the parameters were estimated using the {\tt emcee} package. 
The same procedure was utilized for the single line fits, where only a single slope and y-intercept are necessary. 
A comparison of the goodness-of-fit between the bilinear and linear fits is discussed in Section \ref{gofcomp}. 

\subsection{OCCAM Methodology Changes}\label{methoddiff}
In this work, we use the \citet{cg20} as the starting point of our analysis, which deviates significantly from previous OCCAM studies \citet[][]{myers_2022}\footnote{\natp~used the analysis method described in \citet{occam_p4}, that uses the celestial coordinates, proper motions, and [Fe/H] abundance of stars in the vicinity of the cluster center to determine membership probabilities for these stars to distinguish likely cluster member stars from non-members.
\citet{myers_2022} showed that the \citet{cg20} membership was consistent with the OCCAM proper motion membership \citep[see Figure 2 in][]{myers_2022}.}
All stars included in the \citet{cg20} catalog have been determined to be members of their respective clusters based on the joint probability determined in that work from each star's position, proper motion, and parallax. We refer to these stars as proper motion members for conciseness.
We then perform a cone search in the MWM/APOGEE data for stars within three times the radius containing half of the member stars as calculated by \citet{cg20}, cross-matching with the \citet{cg20} stars, which results in a list of proper motion member stars with MWM/APOGEE data for each cluster in our sample.
Next, we determine radial velocity and [Fe/H] probabilities for those stars using the MWM/APOGEE DR19 data. Due to the more restrictive membership criteria employed in \citet{cg20}, where they retained members with >70\% probability, this analysis uses a $2\sigma$ cut to determine bulk cluster parameters. In practice this is done by keeping stars that have a 0.05 or greater probability of being a cluster member in both radial velocity and [Fe/H] space. 


Due to the increase in the number of clusters in our sample we are able to investigate any variations in the radial metallicity gradient due to differences in the azimuth angle of the cluster with respect to the Galactic center. There are a sufficient number of clusters between $5 \text{ kpc}\leq R_{GC} \leq 14 \text{ kpc}$ and $150\degree \leq \phi \leq 210\degree$ to determine radial metallicity gradients in slices of constant azimuth angle as well as azimuthal gradients for slices of constant \rgc. The results are given in Section \ref{azgrads} and discussed further in Section \ref{azgradsdiss}.

\begin{figure*}
  \epsscale{1.0}
 \plotone{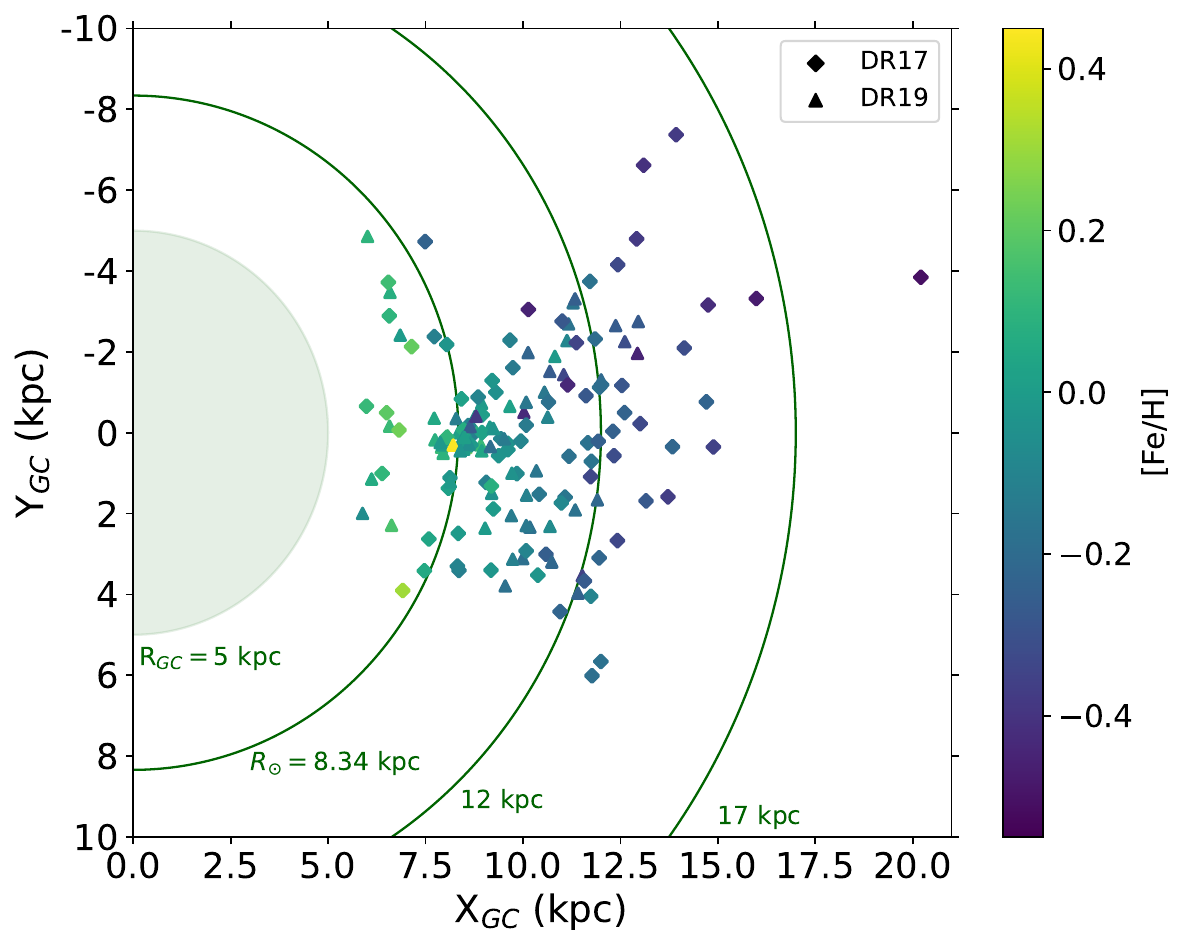}
 	\caption{ \small The OCCAM DR19 sample based on \citet{cg20} membership plotted in the Galactic plane, color-coded by [Fe/H]. Diamond points are clusters that were in the OCCAM DR17 sample, and triangle points are clusters that are new in the DR19 sample. The concentric circles show $R_{GC}$ = 5, 8.34 (the solar circle), 12, and 17 kpc.    }
 	\label{fig:XYfull}
 \end{figure*}

\section{The OCCAM DR19 Sample}\label{dr19}

Our final sample consists of 164 high quality open clusters with 1083 member stars and are shown in Figure \ref{fig:XYfull}. All clusters in the final sample are of sufficient quality to include in our analysis and the visual CMD quality check served to filter out non-clusters and assign a reliability based on the number of stars in the cluster that has MWM/APOGEE data. The clusters in the full sample have basic bulk parameters reported in Table \ref{tab:full_params}, and the detailed chemistry is reported in Table \ref{tab:full_chem}. Both tables are also available as machine-readable tables. 
\noindent
\begin{deluxetable*}{lrrrrrrrrrrrrcr}
\tabletypesize{\tiny}
\tablecaption{Basic Parameters of OCCAM DR19 Clusters \label{tab:full_params}}
	\tablehead{
    \colhead{Cluster} &
    \colhead{l} &
    \colhead{b} &
    \colhead{RA} &
    \colhead{Dec} &
    \colhead{Radius\tablenotemark{a}} &
     \colhead{Age\tablenotemark{a}} &
    \colhead{R$_{GC}$\tablenotemark{a}} & 
    \colhead{R$_{Guide}$\tablenotemark{b}} & 
    \colhead{$\mu_{\alpha}$\tablenotemark{a}} &
    \colhead{$\mu_{\delta}$\tablenotemark{a}} &
    \colhead{RV} & 
    \colhead{[Fe/H]} & 
    \colhead{}&
    \colhead{Num}\\[-4ex] 
    \colhead{name} &
    \colhead{(deg)} &
    \colhead{(deg)} &
    \colhead{(deg)} &
    \colhead{(deg)} &
    \colhead{(deg)} &
     \colhead{(Gyr)} &
    \colhead{(kpc)} &
    \colhead{(kpc)} &
    \colhead{(mas yr$^{-1}$)} &
    \colhead{(mas yr$^{-1}$)} &
    \colhead{(km s$^{-1}$)} & 
    \colhead{(dex)} &
    \colhead{Qual} &
    \colhead{Stars} 
    }
	\startdata
ASCC 123             & 104.672 &  -4.028 & 340.563 & +54.199 &1.29 & 0.045 &  8.40 &  8.81 & $+12.09 \pm  0.47$ & $-1.41 \pm  0.44$ & $ -4.2 \pm 0.02$ & $-0.03 \pm 0.02$ & 1 & 1\\
ASCC 19              & 204.789 & -19.380 & 81.978 & -1.856 &0.60 & 0.010 &  8.64 &  8.91 & $+1.15 \pm  0.25$ & $-1.23 \pm  0.22$ & $+23.2 \pm 0.58$ & $+0.03 \pm 0.02$ & 2 & 2\\
ASCC 21              & 199.883 & -16.530 & 82.212 & +3.607 &0.41 & 0.009 &  8.65 &  8.79 & $+1.40 \pm  0.26$ & $-0.63 \pm  0.24$ & $+41.0 \pm 0.05$ & $+0.09 \pm 0.02$ & 1 & 1\\
ASCC 99              & 16.165 &  -7.682 & 282.172 & -18.384 &0.64 & 0.355 &  8.07 &  8.45 & $+5.19 \pm  0.35$ & $-1.31 \pm  0.46$ & $-29.9 \pm 0.04$ & $+0.15 \pm 0.02$ & 1 & 1\\
Alessi 19            & 39.932 & +12.720 & 274.534 & +11.897 &0.59 & 0.026 &  7.92 &  7.93 & $-1.00 \pm  0.22$ & $-7.07 \pm  0.18$ & $ -7.8 \pm 0.02$ & $+0.13 \pm 0.02$ & 1 & 1\\
Alessi 2             & 152.327 &  +6.336 & 71.557 & +55.184 &0.55 & 0.355 &  8.92 &  9.51 & $-0.90 \pm  0.14$ & $-1.05 \pm  0.12$ & $-10.1 \pm 0.15$ & $+0.12 \pm 0.07$ & 2 & 2\\
Alessi 20            & 117.643 &  -3.695 & 2.643 & +58.757 &0.23 & 0.009 &  8.54 &  8.71 & $+8.20 \pm  0.29$ & $-2.34 \pm  0.28$ & $ -7.5 \pm 0.02$ & $+0.24 \pm 0.02$ & 1 & 1\\
Alessi 21            & 223.373 &  -0.079 & 107.592 & -9.318 &0.47 & 0.066 &  8.78 &  8.86 & $-5.47 \pm  0.19$ & $+2.60 \pm  0.16$ & $+40.0 \pm 0.02$ & $+0.00 \pm 0.02$ & 1 & 1\\
Alessi 62            & 52.799 &  +8.742 & 283.963 & +21.607 &0.26 & 0.692 &  7.97 &  8.96 & $+0.24 \pm  0.16$ & $-1.07 \pm  0.18$ & $+13.7 \pm 0.02$ & $+0.09 \pm 0.02$ & 1 & 1\\
BH 211               & 344.947 &  +0.461 & 255.527 & -41.112 &0.08 & 0.427 &  6.52 &  6.97 & $-0.67 \pm  0.16$ & $-1.94 \pm  0.15$ & $-49.0 \pm 0.44$ & $+0.20 \pm 0.02$ & 2 & 2\\
\multicolumn{15}{c}{......}
\enddata
\tablenotetext{a}{Bulk cluster parameters adopted from \citet{cg20}}\vskip-0.07in
\tablenotetext{b}{Calculated with distances from \citet{cg20}, computed with a solar radius of $R_{\odot} = 8.34$ kpc.}\vskip-0.07in
\tablenotetext{}{(This table is available in its entirety in machine-readable form.)}
\end{deluxetable*}


\begin{deluxetable*}{lrrrrrrrr}
\tabletypesize{\ssmall}
\tablecaption{OCCAM DR19 Sample - Detailed Chemistry \label{tab:full_chem}}
	\tablehead{
    \colhead{Cluster} & 
    \colhead{[Fe/H]} &
    \colhead{[O/H]} & 
    \colhead{[Na/H]} & 
    \colhead{[Mg/H]} & 
    \colhead{[Al/H]} & 
    \colhead{[Si/H]} & 
    \colhead{[S/H]} &
    \colhead{[K/H]} \\[-4ex]
     \colhead{name} &
     \colhead{(dex)} &
     \colhead{(dex)} &
     \colhead{(dex)} &
     \colhead{(dex)} &
     \colhead{(dex)} &
     \colhead{(dex)} &
     \colhead{(dex)} &
     \colhead{(dex)} \\[-2ex]
    \colhead{} &
    \colhead{[Ca/H]} &
    \colhead{[Ti/H]} &
    \colhead{[Cr/H]} & 
    \colhead{[Mn/H]} & 
    \colhead{[Co/H]} & 
    \colhead{[Ni/H]} & 
    \colhead{[Ce/H]} &
    \colhead{[Nd/H]}\\[-4ex] 
     \colhead{} &
     \colhead{(dex)} &
     \colhead{(dex)} &
     \colhead{(dex)} &
     \colhead{(dex)} &
     \colhead{(dex)} &
     \colhead{(dex)} &
     \colhead{(dex)} &
     \colhead{(dex)}
    }
\startdata
ASCC 123           & $-0.03 \pm 0.02$ & $+0.06 \pm 0.02$ & $+0.30 \pm 0.02$ & $-0.11 \pm 0.03$ & $-0.01 \pm 0.05$ & $-0.08 \pm 0.02$ & $-0.00 \pm 0.02$ & $+0.01 \pm 0.02$\\
                   & $-0.01 \pm 0.03$ & $+0.23 \pm 0.03$ & $+0.06 \pm 0.02$ & $-0.15 \pm 0.02$ & $+0.28 \pm 0.13$ & $-0.06 \pm 0.02$ & $+0.14 \pm 0.05$ & $-0.14 \pm 0.05$\\
ASCC 19            & $+0.03 \pm 0.02$ & $+0.02 \pm 0.04$ & $+0.20 \pm 0.38$ & $-0.13 \pm 0.06$ & $-0.11 \pm 0.05$ & $+0.02 \pm 0.02$ & $+0.03 \pm 0.04$ & $+0.08 \pm 0.13$\\
                   & $+0.08 \pm 0.05$ & $+0.36 \pm 0.28$ & $+0.21 \pm 0.17$ & $-0.06 \pm 0.06$ & $-0.96 \pm 1.23$ & $-0.03 \pm 0.02$ & $+0.20 \pm 0.14$ & $-0.43 \pm 0.52$\\
ASCC 21            & $+0.09 \pm 0.02$ & $+0.12 \pm 0.02$ & $-2.36 \pm 0.02$ & $-0.05 \pm 0.02$ & $+0.16 \pm 0.05$ & $-0.03 \pm 0.02$ & $-0.38 \pm 0.02$ & $-0.13 \pm 0.02$\\
                   & $+0.16 \pm 0.02$ & $-0.20 \pm 0.02$ & $+0.02 \pm 0.02$ & $+0.18 \pm 0.02$ & $+0.25 \pm 0.13$ & $+0.05 \pm 0.02$ & $+0.37 \pm 0.05$ & $-0.07 \pm 0.05$\\
ASCC 99            & $+0.15 \pm 0.02$ & $+0.07 \pm 0.02$ & $+0.06 \pm 0.02$ & $+0.06 \pm 0.02$ & $+0.15 \pm 0.05$ & $+0.16 \pm 0.02$ & $+0.14 \pm 0.02$ & $+0.07 \pm 0.02$\\
                   & $+0.13 \pm 0.02$ & $+0.15 \pm 0.02$ & $+0.08 \pm 0.02$ & $+0.17 \pm 0.02$ & $+0.65 \pm 0.13$ & $+0.13 \pm 0.02$ & $+0.46 \pm 0.05$ & $+0.29 \pm 0.05$\\
Alessi 19          & $+0.13 \pm 0.02$ & $+0.17 \pm 0.03$ & $-2.36 \pm 0.04$ & $+0.04 \pm 0.04$ & $+0.16 \pm 0.05$ & $+0.10 \pm 0.03$ & $+0.02 \pm 0.02$ & $+0.04 \pm 0.02$\\
                   & $+0.31 \pm 0.04$ & $+0.55 \pm 0.04$ & $-0.78 \pm 0.03$ & $+0.23 \pm 0.02$ & $+0.58 \pm 0.14$ & $+0.13 \pm 0.02$ & $+0.51 \pm 0.05$ & $+0.14 \pm 0.05$\\
Alessi 2           & $+0.12 \pm 0.07$ & $+0.09 \pm 0.15$ & $-0.99 \pm 1.37$ & $+0.04 \pm 0.07$ & $+0.05 \pm 0.11$ & $+0.03 \pm 0.11$ & $+0.07 \pm 0.23$ & $-1.58 \pm 0.82$\\
                   & $+0.15 \pm 0.17$ & $+0.21 \pm 0.72$ & $+0.33 \pm 0.46$ & $+0.25 \pm 0.04$ & $+0.02 \pm 0.47$ & $+0.09 \pm 0.07$ & $-0.24 \pm 0.50$ & $+0.28 \pm 0.06$\\
Alessi 20          & $+0.24 \pm 0.02$ & $+0.44 \pm 0.02$ & $+0.84 \pm 0.02$ & $+0.17 \pm 0.02$ & $-0.25 \pm 0.05$ & $+0.20 \pm 0.02$ & $+0.60 \pm 0.02$ & $+0.15 \pm 0.02$\\
                   & $+0.28 \pm 0.02$ & $+0.99 \pm 0.02$ & $+0.10 \pm 0.02$ & $+0.12 \pm 0.02$ & $-0.28 \pm 0.13$ & $+0.29 \pm 0.02$ & $-0.55 \pm 0.05$ & $+0.15 \pm 0.05$\\
Alessi 21          & $+0.00 \pm 0.02$ & $+0.12 \pm 0.04$ & $+0.67 \pm 0.02$ & $-0.02 \pm 0.07$ & $+0.06 \pm 0.06$ & $+0.07 \pm 0.05$ & $-0.06 \pm 0.03$ & $-0.09 \pm 0.02$\\
                   & $-0.08 \pm 0.06$ & $-0.69 \pm 0.08$ & $+0.05 \pm 0.03$ & $+0.02 \pm 0.02$ & $+1.25 \pm 0.14$ & $+0.01 \pm 0.02$ & $-0.32 \pm 0.05$ & $-0.09 \pm 0.06$\\
Alessi 62          & $+0.09 \pm 0.02$ & $+0.22 \pm 0.04$ & $-2.36 \pm 0.09$ & $+0.07 \pm 0.07$ & $-0.03 \pm 0.06$ & $+0.06 \pm 0.05$ & $+0.28 \pm 0.02$ & $+0.38 \pm 0.02$\\
                   & $+0.14 \pm 0.06$ & $-0.63 \pm 0.09$ & $-0.54 \pm 0.04$ & $+0.16 \pm 0.02$ & $+0.52 \pm 0.14$ & $+0.12 \pm 0.02$ & $-0.39 \pm 0.05$ & $+0.80 \pm 0.06$\\
BH 211             & $+0.20 \pm 0.02$ & $+0.11 \pm 0.02$ & $+0.46 \pm 0.02$ & $+0.10 \pm 0.04$ & $+0.21 \pm 0.05$ & $+0.20 \pm 0.04$ & $+0.21 \pm 0.02$ & $+0.15 \pm 0.09$\\
                   & $+0.11 \pm 0.04$ & $+0.09 \pm 0.03$ & $+0.16 \pm 0.02$ & $+0.29 \pm 0.02$ & $+0.28 \pm 0.13$ & $+0.18 \pm 0.03$ & $+0.26 \pm 0.05$ & $+0.48 \pm 0.05$\\
\multicolumn{9}{c}{......}\\
\enddata
\tablenotetext{}{(This table is available in its entirety in machine-readable form.}
\end{deluxetable*}

\subsection{SDSS Value Added Catalog Data Access}\label{sec:vac}

Two tables, in FITS format, will be released as Value Added Catalogs (VAC) as part of SDSS-V/DR19, occam\_member-DR19.fits table and the occam\_cluster-DR19.fits table.
The occam\_member table contains multiple IDs, coordinates, and parameters (e.g., proper motions, RVs, and metallicities) for each proper motion member utilized in this work. In addition, the RV membership probabilities and [Fe/H] probabilities determined in the OCCAM pipeline are also reported for each member.  All columns included in the occam\_member VAC file are shown in Table \ref{tab:vac}. 
The occam\_cluster table contains bulk chemistry, motions, and orbital parameters for the 164 open clusters. In a break with previous OCCAM papers, we used stars that are within $2\sigma$ of the cluster mean in proper motion, radial velocity, and [Fe/H] space to determine the bulk cluster parameters. Due to the change in methodology, we no longer have large numbers of background field stars considered in the analysis. This results in Gaussian fits that are wider than previous OCCAM papers, and so a $2\sigma$ membership cutoff was deemed more appropriate for this work. 


\begin{deluxetable}{ll}[ht!]
\tablecaption{A summary of the individual star data included in the DR19 OCCAM VAC \label{tab:vac}}
\tabletypesize{\scriptsize}
\tablehead{
    \colhead{Label} & 
    \colhead{Description}
    }
\startdata
Cluster & The associated open cluster \\
SDSS\_ID \tablenotemark{a}&   MWM {star} ID \\
GaiaDR3\_ID\tablenotemark{b} & \gaia DR3 star ID \\
GaiaDR2\_ID\tablenotemark{c} & \gaia DR2 star ID \\ 
OBJ\_ID\tablenotemark{a} & DR17 APOGEE ID \\
GLON &  Galactic longitude \\
GLAT &  Galactic latitude \\
RAdeg &  right ascension \\
DEdeg & declination \\
V\_RAD \tablenotemark{a}& radial velocity \\ 
E\_V\_RAD\tablenotemark{a} & standard error in V\_RAD \\
STD\_V\_RAD\tablenotemark{a} & $1\sigma$ scatter in V\_RAD \\
PMRA\tablenotemark{b} &  proper motion in right ascension\\
E\_PMRA\tablenotemark{b}  & uncertainty in PMRA \\
PMDE\tablenotemark{b}  & proper motion in declination \\
E\_PMDE\tablenotemark{b}  & uncertainty in PMDEC \\
FeH\_ASPCAP\tablenotemark{a} & [Fe/H] from ASPCAP \\ 
E\_FeH\_ASPCAP\tablenotemark{a} & $1\sigma$ [Fe/H] dispersion \\
CG\_PROB &  membership probability from\\ 
& \citet{cg20} \\
RV\_PROB &  membership probability based \\ &on RV (This study)\\
FEH\_PROB &  membership probability based \\ &on FE\_H\_ASPCAP (This study)\\
EH\_PROB &  membership probability from \\ & \citet{HUNT_2023} \\
XMatch & crossmatch to other open cluster surveys \\[1ex]
\enddata
\tablenotetext{a}{Taken directly from MWM DR19.}\vskip-0.07in
\tablenotetext{b}{From \gaia DR3.} \vskip-0.07in
\tablenotetext{c}{From \gaia DR2}
\end{deluxetable}

\section{Results} \label{sec:results}

\subsection{The Galactic Metallicity Gradient}\label{sec:FeH}

\begin{figure*}[t!]
 	\begin{center}
         \epsscale{1.1}
     \plotone{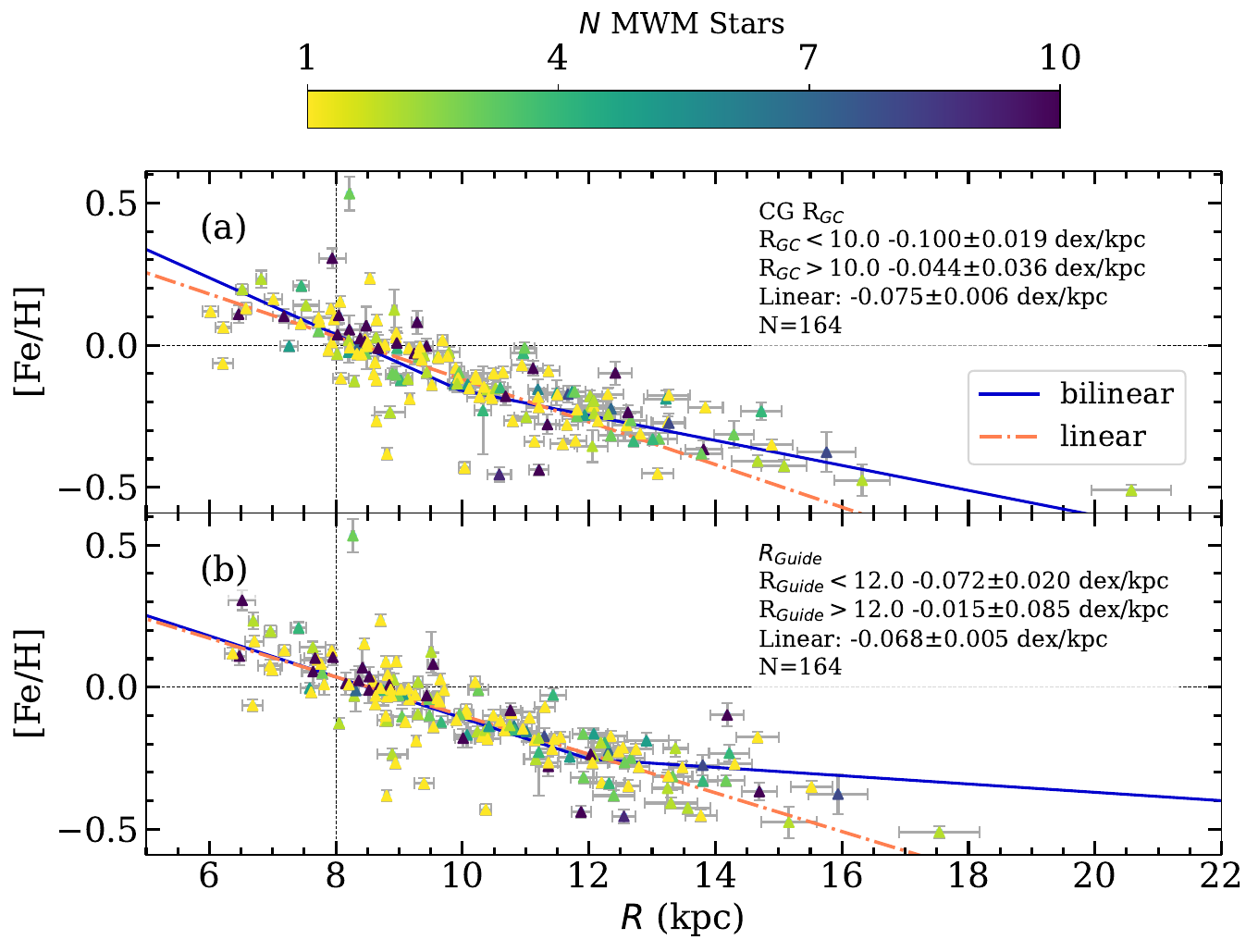} 
 	\end{center}
 	\caption{ \small The Galactic metallicity ([Fe/H]) gradients using the full sample of reliable clusters (shown as triangles), as a function of current Galactocentric radius ($R_{GC}$; top panel (a)) and guiding center radius ($R_{Guide}$; bottom panel (b)). The bilinear fit (blue lines) and linear fit (coral dot-dashed line), are shown. Fit parameters and knee locations are indicated within each panel. The color bar indicates the number of OCCAM member stars in each cluster, saturating at a value of 10 stars. }
 	\label{fig:feh_grad}
 \end{figure*}

With the addition of the MWM/APOGEE DR19 data, the number of open clusters we can use to fit the Galactic metallicity gradient has nearly doubled from 84 clusters in \natp~to 164 in this work. With this significant increase we are able to reliably characterize radial Galactic abundance gradients for 16 elements, covering multiple element families. Additionally, the increase in open cluster data allows for the OCCAM sample to be used to investigate azimuthal variations in the radial [Fe/H] gradient and to characterize the azimuthal gradient for four Galactocentric radius slices for the first time with the OCCAM data.
The [Fe/H] abundance of the OCCAM sample as a function of both $R_{GC}$ (top panel) and $R_{Guide}$ (bottom panel) is shown in Figure \ref{fig:feh_grad}. We use a single linear gradient fit and a two-function linear gradient fit according to the procedure described in \citet{occam_p4}. In the two-function  bilinear case, the ``knee", where the two lines intersect, is allowed to be a free parameter. The fitting procedure takes into account both x and y errors, a 5\% uncertainty in the distance to the cluster was adopted for the Galactocentric radius errors (from solar distance errors) and the $1\sigma$ dispersion in the cluster [Fe/H] abundance was used for the [Fe/H] errors. 

The inner metallicity gradient concerning both \rgc and \rguide is significantly steeper than the corresponding outer slope. We report an inner slope of $-0.100 \pm 0.019 \text{ dex kpc}^{-1}$ for \rgc with the knee located at $10.0 \pm 1.7$ kpc, and an outer slope of $-0.044 \pm 0.036 \text{ dex kpc}^{-1}$. The overall linear gradient of $-0.075 \pm 0.006 \text{ dex kpc}^{-1}$ as a function of $R_{GC}$. With respect to \rguide, we report the knee at $12.0 \pm 2.7$ kpc with an inner slope of $-0.072 \pm 0.020 \text{ dex kpc}^{-1}$ and an outer slope of $-0.015 \pm 0.085 \text{ dex kpc}^{-1}$. The overall linear gradient with respect to \rguide was determine to be $-0.068 \pm 0.005 \text{ dex kpc}^{-1}$ in this work. 
The inner and outer slopes, along with the knee locations, are recorded in Table \ref{tab:fehgradients}, as well as, the slopes of the single linear function fits, and the number of clusters used in each fit.  \\ 

\begin{deluxetable}{lrrrrr}[h!]
\tabletypesize{\scriptsize}
\tablecaption{OCCAM DR19 {[Fe/H]} Gradients \label{tab:fehgradients}}
	\tablehead{
    \colhead{Selection} &
    \colhead{Type} &
    \colhead{Gradient} &
    \colhead{Knee} &
    \colhead{ N} &
    \colhead{AIC}\\[-2.5ex]
    \colhead{} &
    \colhead{} &
    \colhead{(dex kpc$^{-1}$)} &
    \colhead{(kpc)} &
    \colhead{} &
    \colhead{Score}\\[-4.5ex]
    }
	\startdata
\multicolumn{6}{c}{d[Fe/H]/d$R_{GC}$}\\[0.5ex]\hline
All                 & Linear & $-0.075 \pm 0.006$ & \multicolumn{1}{c}{\nodata} & 164 & 2.54 \\
Inner fit           & Knee  & $-0.100 \pm 0.019$ &   $10.0 \pm 1.7$  & 164 & 4.96\\
Outer fit           & Knee  & $-0.044 \pm 0.036$ &   $10.0 \pm 1.7$  & 164 & 4.96 \\\hline
$Age \le 0.4$       & Linear & $-0.107 \pm 0.012$ & \multicolumn{1}{c}{\nodata} & 52 & \multicolumn{1}{c}{\nodata}\\ 
$0.4 < Age \le 0.8$ & Linear & $-0.074 \pm 0.013$ & \multicolumn{1}{c}{\nodata} & 33 & \multicolumn{1}{c}{\nodata}\\
$0.8 < Age \le 2.0$ & Linear & $-0.058 \pm 0.010$ & \multicolumn{1}{c}{\nodata} & 49 & \multicolumn{1}{c}{\nodata}\\ 
$2.0 < Age$.        & Linear & $-0.087 \pm 0.013$ & \multicolumn{1}{c}{\nodata} & 30 & \multicolumn{1}{c}{\nodata}\\\hline 
 \multicolumn{5}{c}{{d[Fe/H]/d$R_{Guide}$}}\\[0.5ex]\hline
All                 & Linear & $-0.068 \pm 0.005$ & \multicolumn{1}{c}{\nodata} & 164 & 2.95\\
Inner fit           & Knee  & $-0.072 \pm 0.020$ &   $12.0 \pm 2.7$  & 164 & 6.89\\
Outer fit           & Knee  & $-0.015 \pm 0.085$ &   $12.0 \pm 2.7$  & 164  & 6.89\\\hline
$Age \le 0.4$       & Linear & $-0.088 \pm 0.010$ & \multicolumn{1}{c}{\nodata} & 52 & \multicolumn{1}{c}{\nodata}\\ 
$0.4 < Age \le 0.8$ & Linear & $-0.067 \pm 0.013$ & \multicolumn{1}{c}{\nodata} & 33 & \multicolumn{1}{c}{\nodata}\\
$0.8 < Age \le 2.0$ & Linear & $-0.053 \pm 0.010$ & \multicolumn{1}{c}{\nodata} & 49 & \multicolumn{1}{c}{\nodata}\\ 
$2.0 < Age$         & Linear & $-0.084 \pm 0.013$ & \multicolumn{1}{c}{\nodata} & 30 & \multicolumn{1}{c}{\nodata}
\enddata

\end{deluxetable}

 \subsubsection{Comparing the Linear and Bilinear Fits}\label{gofcomp}

To assess whether the overall Galactic metallicity ([Fe/H]) gradient prefers a linear or bilinear fit, we compute the Akaike information criterion (AIC). The AIC is calculated according to the formula:
\begin{equation}
    \text{AIC} = 2k - 2ln(\hat{L})
\end{equation}
where $k$ is the number of free parameters in the model and $\hat{L}$ is the maximized value of the likelihood function. 
This metric has the advantage of evaluating goodness-of-fit while penalizing the addition of free parameters to the model. 
According to the AIC, the metallicity gradient as a function of both \rgc and \rguide prefers linear fits (shown as the coral dot-dashed line in Figure \ref{fig:feh_grad}) over a bilinear fit. The AIC scores for both the linear and bilinear fits are given in Table \ref{tab:fehgradients}. 
The added complexity of the bilinear fit does not improve the goodness-of-fit by a large enough margin to outweigh the penalty incurred by doubling the number of free parameters from two to four.   
For the remainder of this work we fit linear trends to the data.

\noindent
\begin{deluxetable*}{lcrcrcrcrcrcr}
\tabletypesize{\tiny}
\tablecaption{OCCAM DR19 Abundance Gradients \label{tab:gradients}}
	\tablehead{
    \colhead{Age range} &
    \colhead{All} &
    \colhead{N} &
    \colhead{All} &
    \colhead{N} &
    \colhead{$Age \le 0.4$} &
    \colhead{N} &
    \colhead{$0.4 < Age \le 0.8$} &
    \colhead{N} &
    \colhead{$0.8 < Age \le 2.0$} &
    \colhead{N} &
     \colhead{$2.0 < Age$} & 
    \colhead{N} \\[-4.5ex]
    \colhead{$R$ range} &
    \colhead{All $R$} &
    \colhead{} &
    \colhead{$R < 14$} &
    \colhead{} &
    \colhead{All $R$} &
    \colhead{} &
    \colhead{All $R$} &
    \colhead{} &
    \colhead{All $R$} &
    \colhead{} &
     \colhead{All $R$} & 
    \colhead{} \\[-4.5ex]
    \colhead{Gradient} &
    \colhead{(dex kpc$^{-1}$)} &
    \colhead{} &
    \colhead{(dex kpc$^{-1}$)} &
    \colhead{} &
    \colhead{(dex kpc$^{-1}$)} &
    \colhead{} &
    \colhead{(dex kpc$^{-1}$)} &
    \colhead{} &
    \colhead{(dex kpc$^{-1}$)} &
    \colhead{} &
     \colhead{(dex kpc$^{-1}$)} & 
    \colhead{} \\[-6ex]
    }
    \startdata
\multicolumn{11}{c}{Gradients for $R_{GC}$} \\[1ex] \hline
{d[O/Fe]/d$R_{GC}$} & $-0.000 \pm 0.006$ & 164 & $-0.001 \pm 0.007$ & 156 & $-0.003 \pm 0.013$ & 52 & $+0.004 \pm 0.015$ & 33 & $+0.006 \pm 0.011$ & 49 & $+0.012 \pm 0.015$ & 30\\
{d[Mg/Fe]/d$R_{GC}$} & $+0.010 \pm 0.006$ & 164 & $+0.009 \pm 0.006$ & 156 & $-0.001 \pm 0.013$ & 52 & $+0.005 \pm 0.015$ & 33 & $+0.003 \pm 0.011$ & 49 & $+0.016 \pm 0.015$ & 30\\
{d[Si/Fe]/d$R_{GC}$} & $+0.004 \pm 0.006$ & 164 & $+0.003 \pm 0.007$ & 156 & $-0.001 \pm 0.013$ & 52 & $-0.002 \pm 0.015$ & 33 & $+0.000 \pm 0.011$ & 49 & $+0.002 \pm 0.015$ & 30\\
{d[S/Fe]/d$R_{GC}$} & $+0.014 \pm 0.006$ & 160 & $+0.015 \pm 0.007$ & 152 & $-0.001 \pm 0.013$ & 51 & $+0.024 \pm 0.014$ & 33 & $+0.013 \pm 0.011$ & 47 & $+0.013 \pm 0.016$ & 29\\
{d[Ca/Fe]/d$R_{GC}$} & $-0.004 \pm 0.006$ & 163 & $-0.004 \pm 0.007$ & 155 & $-0.016 \pm 0.014$ & 51 & $+0.001 \pm 0.015$ & 33 & $+0.002 \pm 0.011$ & 49 & $+0.007 \pm 0.015$ & 30\\
{d[Ti/Fe]/d$R_{GC}$} & $-0.016 \pm 0.006$ & 153 & $-0.017 \pm 0.007$ & 145 & $+0.040 \pm 0.014$ & 47 & $+0.029 \pm 0.016$ & 31 & $-0.055 \pm 0.011$ & 47 & $+0.000 \pm 0.018$ & 28\\[1ex] \hline
{d[Cr/Fe]/d$R_{GC}$} & $+0.016 \pm 0.006$ & 153 & $+0.018 \pm 0.007$ & 145 & $+0.061 \pm 0.014$ & 48 & $+0.054 \pm 0.016$ & 30 & $+0.027 \pm 0.011$ & 46 & $-0.006 \pm 0.018$ & 29\\
{d[Mn/Fe]/d$R_{GC}$} & $-0.002 \pm 0.006$ & 162 & $-0.003 \pm 0.006$ & 154 & $+0.006 \pm 0.013$ & 52 & $-0.009 \pm 0.014$ & 33 & $-0.003 \pm 0.011$ & 47 & $-0.011 \pm 0.015$ & 30\\
{d[Co/Fe]/d$R_{GC}$} & $-0.048 \pm 0.010$ & 125 & $-0.057 \pm 0.011$ & 118 & $-0.065 \pm 0.021$ & 42 & $-0.047 \pm 0.025$ & 28 & $-0.038 \pm 0.019$ & 32 & $+0.000 \pm 0.022$ & 23\\
{d[Ni/Fe]/d$R_{GC}$} & $-0.015 \pm 0.006$ & 164 & $-0.017 \pm 0.006$ & 156 & $-0.043 \pm 0.013$ & 52 & $-0.008 \pm 0.014$ & 33 & $-0.007 \pm 0.011$ & 49 & $-0.006 \pm 0.015$ & 30\\[1ex] \hline
{d[Na/Fe]/d$R_{GC}$} & $-0.027 \pm 0.007$ & 138 & $-0.029 \pm 0.007$ & 131 & $-0.041 \pm 0.014$ & 41 & $-0.012 \pm 0.017$ & 30 & $+0.002 \pm 0.012$ & 43 & $-0.004 \pm 0.020$ & 24\\
{d[Al/Fe]/d$R_{GC}$} & $+0.008 \pm 0.007$ & 163 & $+0.007 \pm 0.007$ & 155 & $-0.014 \pm 0.015$ & 52 & $-0.004 \pm 0.016$ & 33 & $+0.012 \pm 0.012$ & 48 & $+0.003 \pm 0.016$ & 30\\
{d[K/Fe]/d$R_{GC}$} & $+0.011 \pm 0.006$ & 158 & $+0.011 \pm 0.007$ & 150 & $-0.030 \pm 0.013$ & 50 & $-0.015 \pm 0.015$ & 31 & $+0.029 \pm 0.011$ & 48 & $+0.009 \pm 0.018$ & 29\\[1ex] \hline
{d[Ce/Fe]/d$R_{GC}$} & $+0.087 \pm 0.007$ & 143 & $+0.093 \pm 0.008$ & 137 & $+0.106 \pm 0.015$ & 47 & $+0.136 \pm 0.017$ & 32 & $+0.045 \pm 0.013$ & 40 & $+0.025 \pm 0.019$ & 24\\
{d[Nd/Fe]/d$R_{GC}$} & $-0.052 \pm 0.008$ & 131 & $-0.055 \pm 0.008$ & 128 & $-0.026 \pm 0.016$ & 47 & $-0.108 \pm 0.017$ & 31 & $-0.049 \pm 0.014$ & 35 & $-0.028 \pm 0.024$ & 18\\[1ex] \hline
\multicolumn{11}{c}{Gradients for $R_{Guide}$}\\[1ex]\hline
{d[O/Fe]/d$R_{Guide}$} & $+0.003 \pm 0.006$ & 164 & $+0.002 \pm 0.006$ & 156 & $+0.013 \pm 0.012$ & 52 & $+0.010 \pm 0.015$ & 33 & $+0.004 \pm 0.011$ & 49 & $+0.011 \pm 0.015$ & 30\\
{d[Mg/Fe]/d$R_{Guide}$} & $+0.007 \pm 0.006$ & 164 & $+0.006 \pm 0.006$ & 156 & $-0.005 \pm 0.012$ & 52 & $+0.008 \pm 0.015$ & 33 & $+0.003 \pm 0.010$ & 49 & $+0.014 \pm 0.014$ & 30\\
{d[Si/Fe]/d$R_{Guide}$} & $+0.003 \pm 0.006$ & 164 & $+0.003 \pm 0.006$ & 156 & $-0.000 \pm 0.012$ & 52 & $-0.000 \pm 0.015$ & 33 & $-0.000 \pm 0.011$ & 49 & $+0.002 \pm 0.015$ & 30\\
{d[S/Fe]/d$R_{Guide}$} & $+0.009 \pm 0.006$ & 160 & $+0.009 \pm 0.006$ & 152 & $-0.016 \pm 0.011$ & 51 & $+0.024 \pm 0.015$ & 33 & $+0.013 \pm 0.011$ & 47 & $+0.012 \pm 0.016$ & 29\\
{d[Ca/Fe]/d$R_{Guide}$} & $-0.006 \pm 0.006$ & 163 & $-0.007 \pm 0.006$ & 155 & $-0.025 \pm 0.012$ & 51 & $+0.006 \pm 0.015$ & 33 & $+0.002 \pm 0.011$ & 49 & $+0.006 \pm 0.015$ & 30\\
{d[Ti/Fe]/d$R_{Guide}$} & $-0.019 \pm 0.006$ & 153 & $-0.020 \pm 0.006$ & 145 & $+0.054 \pm 0.012$ & 47 & $+0.015 \pm 0.016$ & 31 & $-0.064 \pm 0.011$ & 47 & $-0.006 \pm 0.017$ & 28\\[1ex] \hline
{d[Cr/Fe]/d$R_{Guide}$} & $+0.014 \pm 0.006$ & 153 & $+0.016 \pm 0.007$ & 145 & $+0.065 \pm 0.012$ & 48 & $+0.027 \pm 0.016$ & 30 & $+0.035 \pm 0.012$ & 46 & $-0.004 \pm 0.017$ & 29\\
{d[Mn/Fe]/d$R_{Guide}$} & $-0.002 \pm 0.006$ & 162 & $-0.003 \pm 0.006$ & 154 & $-0.001 \pm 0.011$ & 52 & $-0.002 \pm 0.014$ & 33 & $-0.003 \pm 0.011$ & 47 & $-0.011 \pm 0.014$ & 30\\
{d[Co/Fe]/d$R_{Guide}$} & $-0.055 \pm 0.010$ & 125 & $-0.061 \pm 0.011$ & 118 & $-0.063 \pm 0.020$ & 42 & $-0.054 \pm 0.023$ & 28 & $-0.046 \pm 0.019$ & 32 & $-0.010 \pm 0.023$ & 23\\
{d[Ni/Fe]/d$R_{Guide}$} & $-0.016 \pm 0.006$ & 164 & $-0.017 \pm 0.006$ & 156 & $-0.042 \pm 0.011$ & 52 & $-0.007 \pm 0.015$ & 33 & $-0.006 \pm 0.010$ & 49 & $-0.007 \pm 0.014$ & 30\\[1ex] \hline
{d[Na/Fe]/d$R_{Guide}$} & $-0.028 \pm 0.007$ & 138 & $-0.029 \pm 0.007$ & 131 & $-0.034 \pm 0.013$ & 41 & $-0.022 \pm 0.016$ & 30 & $+0.000 \pm 0.012$ & 43 & $-0.006 \pm 0.019$ & 24\\
{d[Al/Fe]/d$R_{Guide}$} & $+0.006 \pm 0.007$ & 163 & $+0.005 \pm 0.007$ & 155 & $-0.019 \pm 0.013$ & 52 & $+0.000 \pm 0.016$ & 33 & $+0.012 \pm 0.012$ & 48 & $+0.003 \pm 0.016$ & 30\\
{d[K/Fe]/d$R_{Guide}$} & $-0.006 \pm 0.006$ & 158 & $-0.007 \pm 0.006$ & 150 & $-0.093 \pm 0.012$ & 50 & $+0.006 \pm 0.015$ & 31 & $+0.028 \pm 0.011$ & 48 & $+0.006 \pm 0.017$ & 29\\[1ex] \hline
{d[Ce/Fe]/d$R_{Guide}$} & $+0.084 \pm 0.007$ & 143 & $+0.088 \pm 0.007$ & 137 & $+0.096 \pm 0.014$ & 47 & $+0.107 \pm 0.017$ & 32 & $+0.048 \pm 0.014$ & 40 & $+0.040 \pm 0.019$ & 24\\
{d[Nd/Fe]/d$R_{Guide}$} & $-0.046 \pm 0.008$ & 131 & $-0.047 \pm 0.008$ & 128 & $-0.013 \pm 0.014$ & 47 & $-0.085 \pm 0.018$ & 31 & $-0.064 \pm 0.014$ & 35 & $-0.028 \pm 0.023$ & 18\\[1ex]
\enddata
\end{deluxetable*}

\subsection{Galactic Trends for Other Elements}\label{all_elems}

\subsubsection{$\alpha-$Elements -- O, Mg, Si, S, Ca, Ti}\label{alpha}

\begin{figure}
    \epsscale{1.1}
 	\plotone{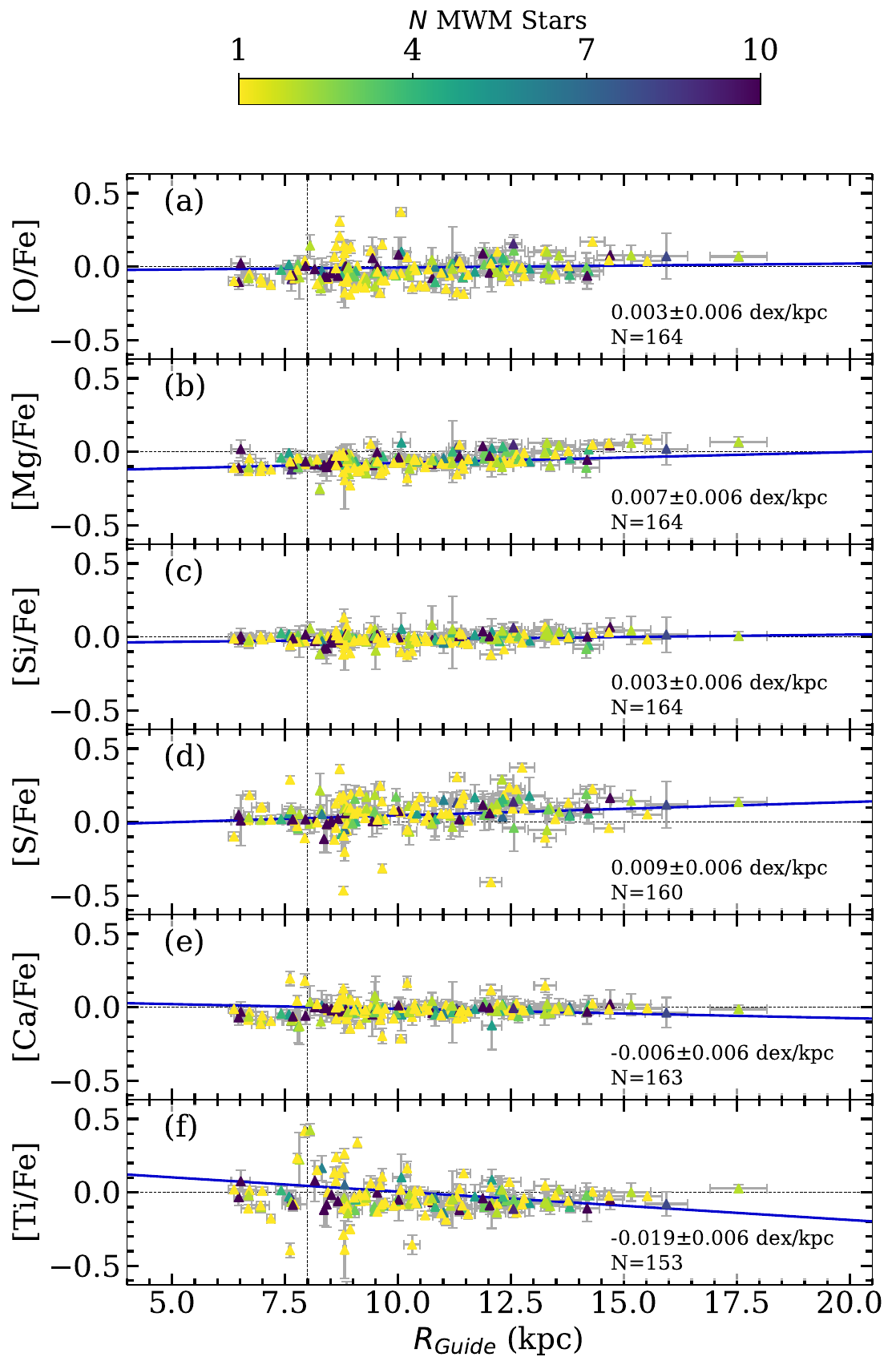}
 	\caption{ \small The [X/Fe] versus \rguide trend for the $\alpha$-elements (O, Mg, Si, S, Ca, Ti). As in Figure \ref{fig:feh_grad}, the color bar indicates the number of member stars, saturating at 10. Clusters with a $1\sigma$ scatter higher than 0.2 dex for a specific element were not used to determine the fit (solid blue line) and are not plotted. The derived gradient and number of clusters (N) are shown in each panel.}
 	\label{fig:alpha}
 \end{figure}

We show the [$\alpha$/Fe] abundance gradients for six $\alpha$-elements (O, Mg, Si, S, Ca, Ti) with respect to \rguide in Figure \ref{fig:alpha}. We find shallow positive slopes for Mg and S. 
O, Si, and Ca all display gradients consistent with no trend.  
In Ti, we find a negative slope, in contrast to the flat or slightly positive gradients the rest of the alpha elements display. 
We do note that the Ti abundances show a larger scatter than the other elements considered here. Ti abundances consistently exhibit larger scatter in APOGEE spectra, regardless of the abundance determination method employed (\citealt{Souto2019}, \citealt{Jonsson2020}).
The gradients determined with respect to \rgc and \rguide are in good agreement with each other, with no significant differences. 

\subsubsection{Iron-Peak Elements -- Cr, Mn, Co, Ni}\label{ironpeak}

\begin{figure}[t!]
    \begin{center}
    \epsscale{1.1}
 	\plotone{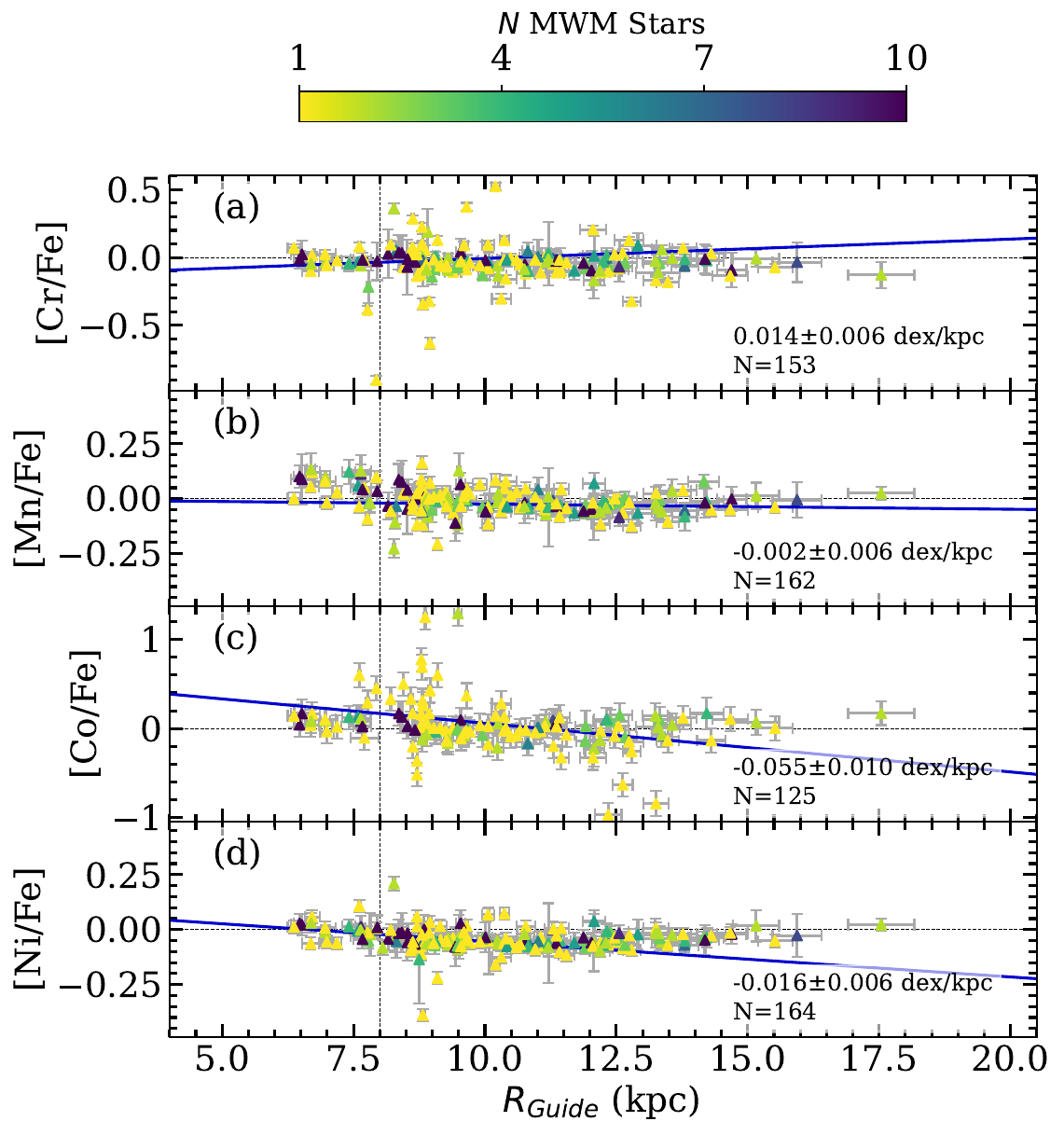}
 	\vskip0.1in
 	\caption{ \small Same as Figure \ref{fig:alpha}, but for the iron-peak elements (Cr, Mn, Co, Ni).}
 	\label{fig:ironpeak}
 	\end{center}
\end{figure}

[X/Fe] abundance gradients for four iron-peak elements (Cr, Mn, Co, Ni)\footnote{While abundance measurements for V and Cu exist within SDSS-V/MWM, these elements were not reliably measured and are therefore left out of our analysis. See \sm et al., {\em in press} for further details.} with respect to \rguide are shown in Figure \ref{fig:ironpeak}. 
Cr shows a shallow positive trend with respect to \rguide. The slope for Mn is consistent with no trend and Ni has a shallow negative trend. 
Co shows a negative trend but we note there is significant scatter in Co, with two clusters near $-1$ dex, Haffner 4 ([Co/Fe] $= -0.97$ dex) and NGC 2423 ([Co/Fe] = $-1.01$ dex), and two clusters over $+1$ dex, Alessi 21 ([Co/Fe] = $1.25$ dex) and Czenik 18 ([Co/Fe] = $1.29$ dex). 
Co abundances within ASPCAP in DR19 show temperature dependence within open clusters (\sm et al, {\em in press}, this problem was present in the DR17 ASPCAP results as well \citep[][]{holtzman_2018}. 
Abundance gradients for the iron-peak elements with respect to \rgc were also determined and recorded with the \rguide gradients in Table \ref{tab:gradients}. There are no significant differences between the gradients determined using \rguide and those determined using \rgc. 

\subsubsection{Odd-Z Elements -- Na, Al, K}

\begin{figure}
    \epsscale{1.1}
 	\plotone{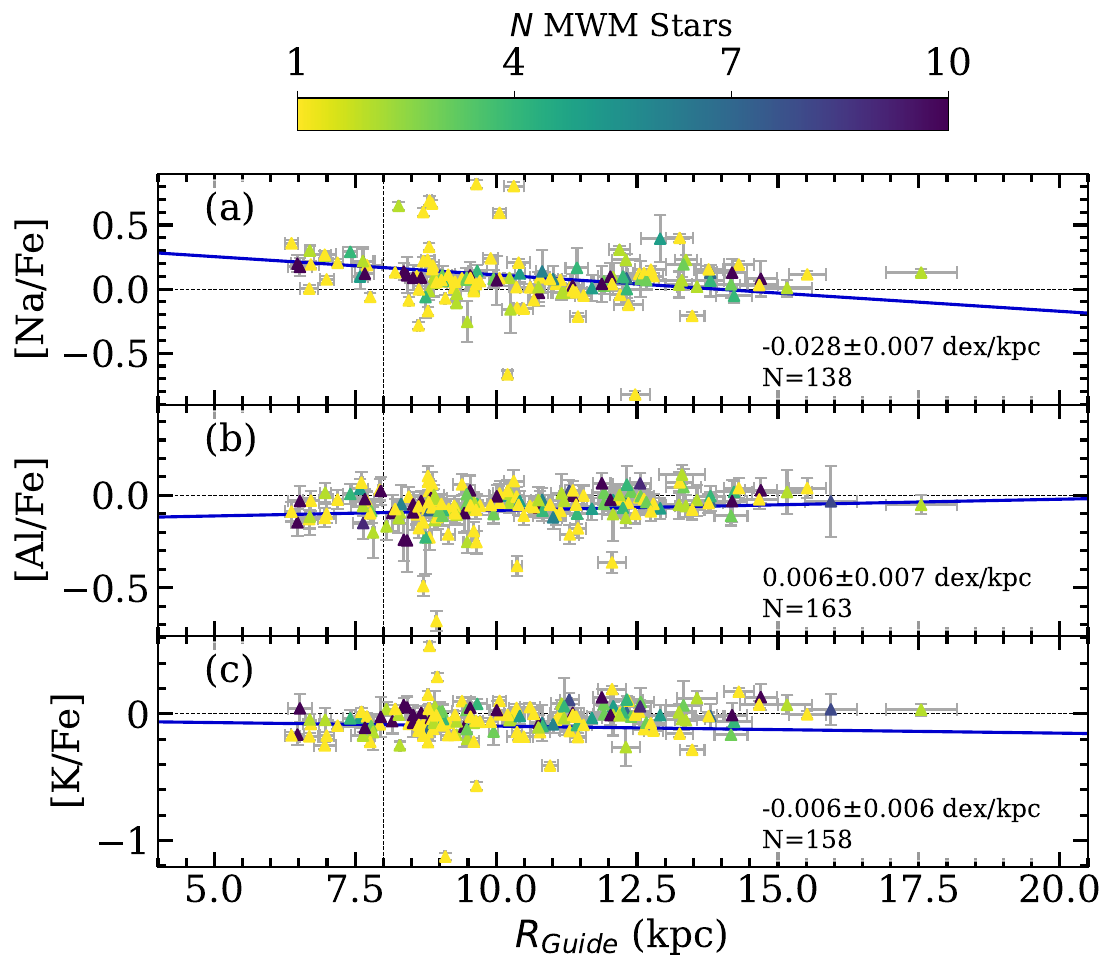}
 	\caption{ \small Same as Figure \ref{fig:alpha} but for the ``odd-z" elements (Na, Al, K). }
 	\label{fig:oddz}
\end{figure}

In Figure \ref{fig:oddz}, the [X/Fe] abundance trends for three odd-z elements (Na, Al and K) are shown. Al and K both show gradients that are consistent with flat. Al is slightly positive and K slightly negative, but both show no trend within the error bars. Na shows a negative gradient but has the most scatter among the three elements. 
K has a significant outlier, IC 348 ([K/Fe] $= -1.13$ dex), which only has data for a single star in the cluster. We report both these gradients as well as those determined with respect to \rgc in Table \ref{tab:gradients}. We note that the K gradient with respect to \rgc shows a shallow positive slope of $+0.011 \pm 0.006 \text{ dex kpc}^{-1}$ in constrast with the flat slope with respect to \rguide.

\subsubsection{Neutron Capture Elements -- Ce, Nd}\label{neutron}

\begin{figure}[t!]
    \begin{center}
    \epsscale{1.2}
 	\plotone{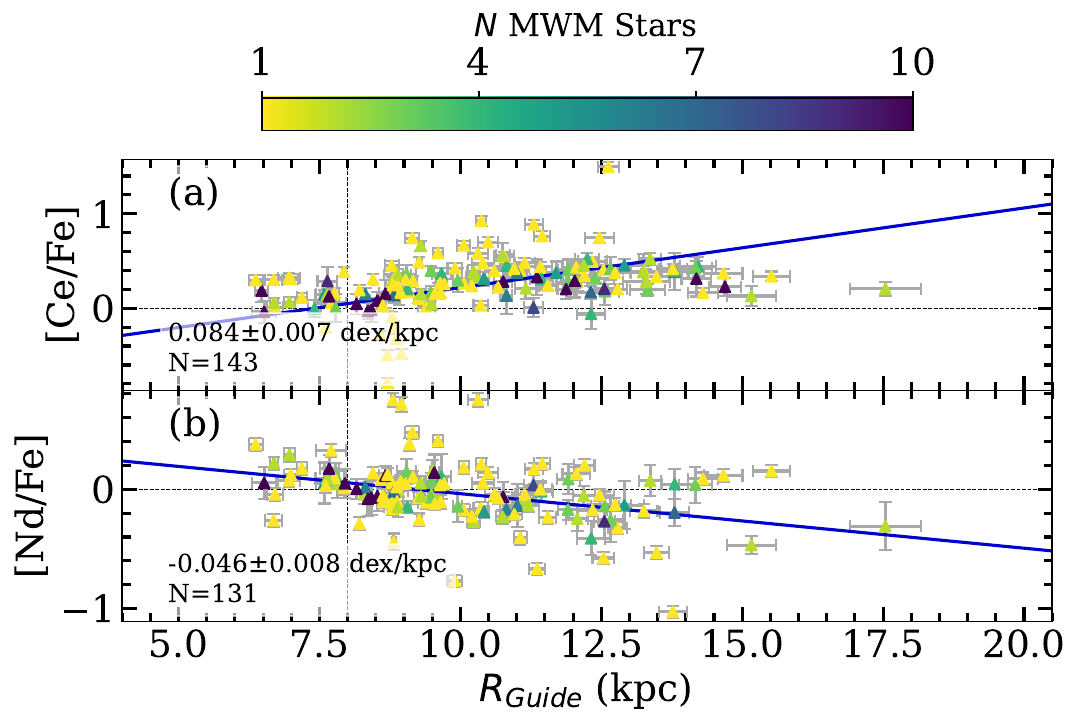}
 	\vskip0.1in
 	\caption{ \small Same as Figure \ref{fig:alpha} but for the neutron-capture elements cerium (Ce) and neodymium (Nd). }
 	\label{fig:neutron}
 	\end{center}
\end{figure}

We can now report the [Ce/Fe] abundance trend using 143 clusters, and for the first time we can characterize the [Nd/Fe] abundance trend using 131 open clusters. We report a positive slope of $0.084 \pm 0.007 \text{ dex kpc}^{-1}$ for [Ce/Fe] with respect to \rguide and a negative slope of $-0.046 \pm 0.008 \text{ dex kpc}^{-1}$ with respect to \rguide for [Nd/Fe]. We note the relatively large scatter in both elements, but Nd in particular has significant scatter. As with the other elements analyzed, we determine the gradients with respect to \rgc as well and see good agreement with the gradients shown in Figure \ref{fig:neutron}. All slopes are recorded in Table \ref{tab:gradients}. 

We report the trends as found in this work, however, we note that both Ce and Nd are measured from weak lines (\citealt{cunha_2017} and \citealt{Hasselquist_2016}) in MWM/DR19 and have relatively high scatter.
They were found to be of "Fair" quality according to an internal quality analysis, the same category as Na, Ti, and Co. For a complete discussion on the quality of the abundance measurements in MWM/DR19, see \citet{Szabolcs_2025}.

\subsection{Azimuthal Gradients}\label{azgrads}

\begin{figure}
  \epsscale{1.2}
 \plotone{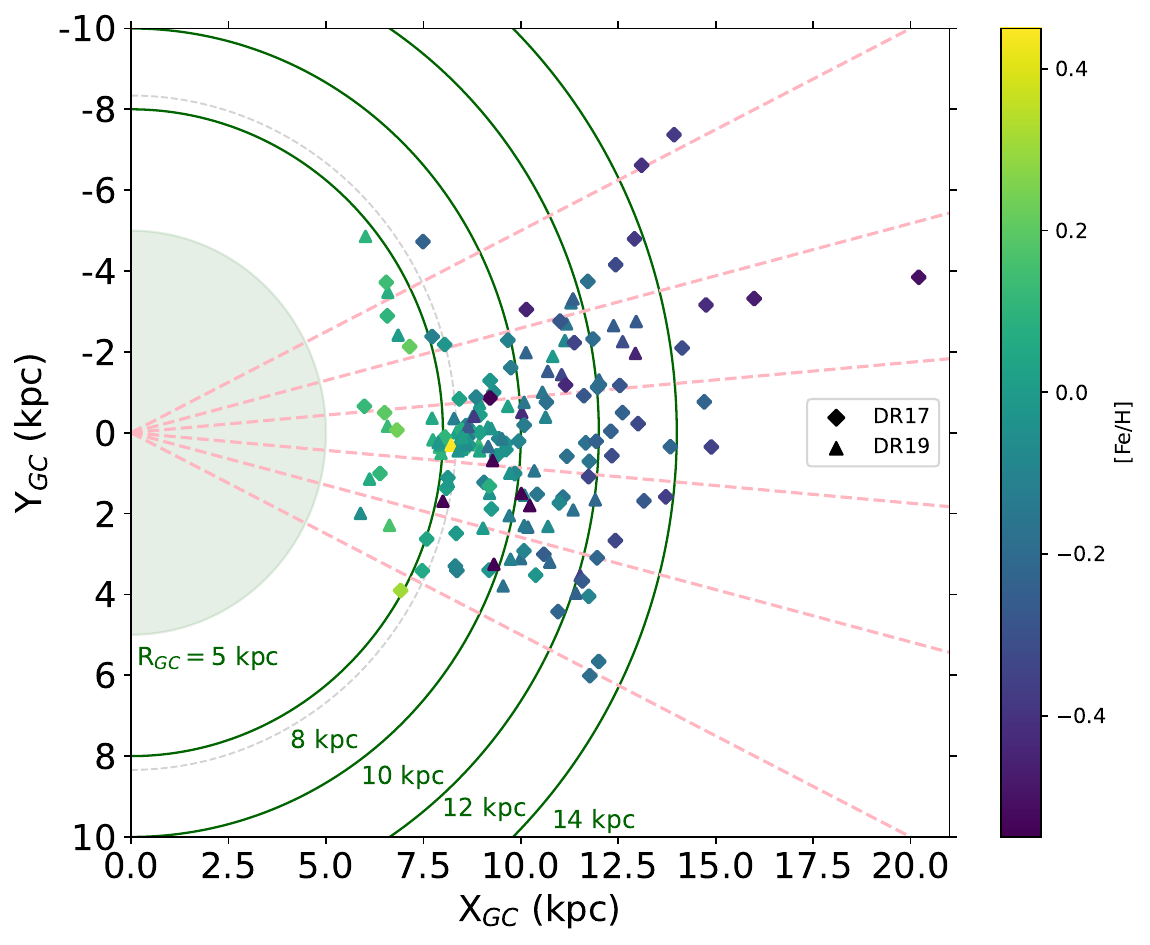}
 	\caption{ \small Same plot as Figure \ref{fig:XYfull}, where the entire cluster sample is plotted in X-Y space, colored by the mean [Fe/H] abundance. The diamond points denote clusters that were in OCCAM DR17/\citet{myers_2022}, while triangle points denote clusters new to this DR19 analysis. The solid green lines (\rgc = 5,8,10,12,14 kpc) delineate the annuli used for analyzing azimuthal gradients (see Section \ref{azgrads} and Figure \ref{fig:az_azimuth}). 
    The pink dashed lines indicate the boundaries of the azimuthal slices used for analyzing radial gradients (see Section \ref{azgrads} and Figure \ref{fig:az_radial}). The grey dashed line shows the solar Galactocentric radius used in this work ($R_{\odot}=8.34$ kpc).}
 	\label{fig:XYaz}
 \end{figure}

With the additional stars observed by SDSS-V/MWM and included in DR19, for the first time, the OCCAM sample has sufficient coverage to investigate the effects of azimuthal variations on the overall radial Galactic [Fe/H] gradient. This has been accomplished in two ways: first by dividing our clusters into five azimuth angle wedges with respect to the Galactic center. These slices are shown in Figure \ref{fig:XYaz} where the pink dashed lines denote the boundaries and the Sun is at an azimuth angle of 180$\degree$. Second, we divide our sample into four \rgc annuli (5--8 kpc, 8--10 kpc, 10--12 kpc, 12--14 kpc) 
and determine the [Fe/H] gradient with respect to the azimuth angle. \rgc was used for this analysis because the X and Y positions of the clusters in Galactocentric coordinates was used to determine the azimuth angle making \rgc the appropriate choice of radius. 
The radius slices are demarcated in Figure \ref{fig:XYaz} by the green lines. While some clusters in the OCCAM sample exist past 14 kpc, there are not sufficient numbers in all azimuth bins to warrant their inclusion. Similarly, the radius bin closest to the Galactic center was extended inward an extra kiloparsec to include several more clusters in the analysis.  

\subsubsection{Radial Gradients in Azimuth Slices}

\begin{figure}[t!]
    \begin{center}
    \epsscale{1.2}
 	\plotone{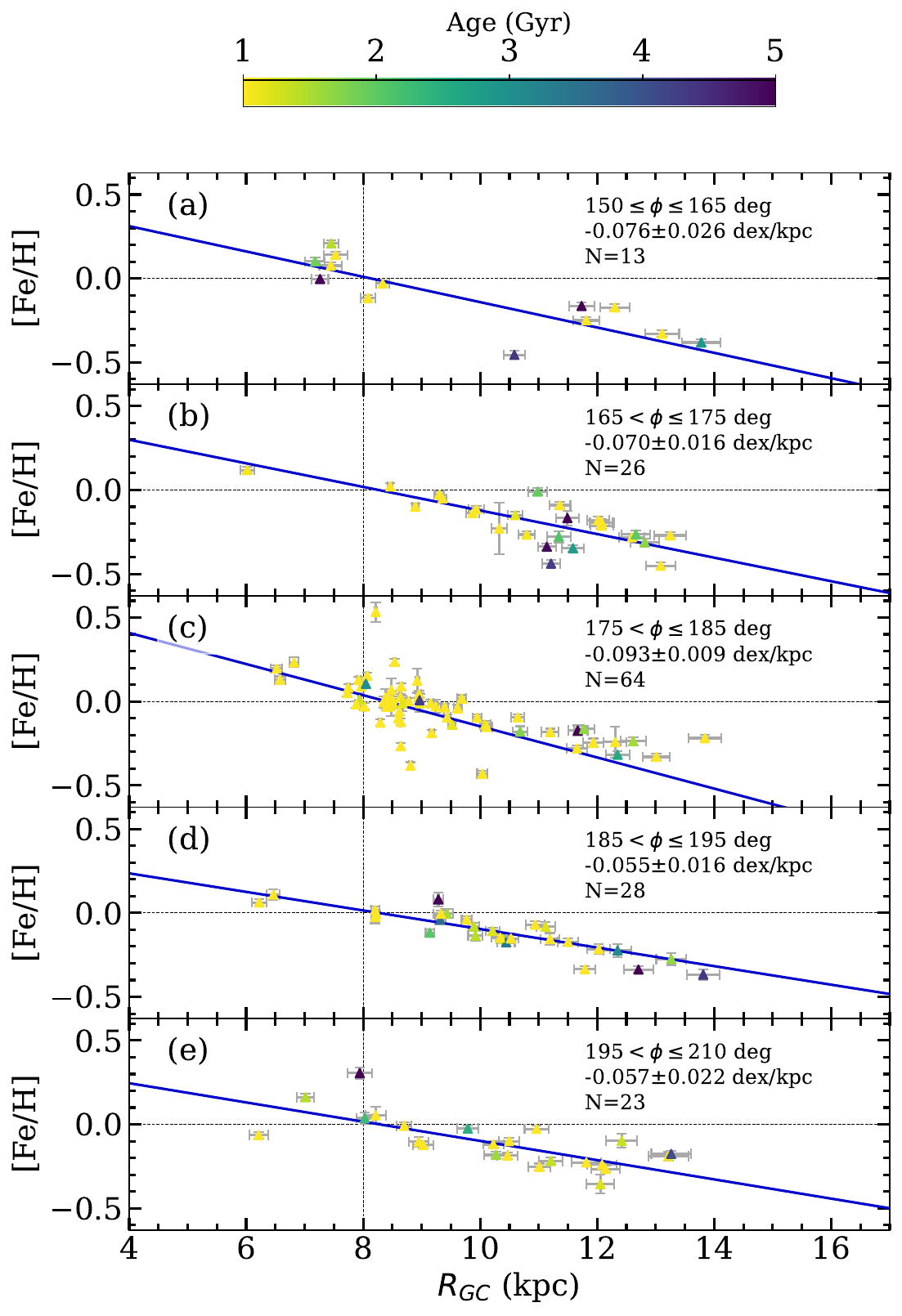}
 	\vskip0.1in
 	\caption{ \small Metallicity gradients ([Fe/H]) as a function of Galactocentric radius ($R_{GC}$) for five slices in azimuth angle ($\phi$): (a) $150\degree \leq \phi \leq 165\degree$, (b) $165\degree < \phi \leq 175\degree$, (c) $175\degree < \phi \leq 185\degree$, (d) $185\degree < \phi \leq 195\degree$, and (e) $195\degree < \phi \leq 210\degree$. Points are colored by the cluster age in Gyr, saturating at younger than 1 Gyr and older than 5 Gyr. The linear fits (blue lines), derived gradients, and number of clusters (N) used are shown in each panel. Fits were determined using only the clusters with $5 \text{ kpc} \leq R_{GC} \leq 14$ kpc. The regions corresponding to these azimuth slices are shown on an X-Y plot in Figure \ref{fig:XYaz}. }
 	\label{fig:az_radial}
 	\end{center}
\end{figure}

We show the radial [Fe/H] gradients as a function of \rgc for clusters between $5-14$ kpc in five azimuth angle bins spanning $\phi = 150 \degree$ to $\phi = 210\degree$ in Figure \ref{fig:az_radial}. All the azimuth slices have negative slopes with varying levels of steepness. Three of the five azimuth slices gradients are within the margin of error of the overall linear slope with respect to \rgc of $-0.075 \pm 0.006 \text{ dex kpc}^{-1}$. However, the bin containing the most clusters, including the solar neighborhood, has a gradient considerably steeper than the overall linear slope and more closely matches the inner slope of the bilinear fit for \rgc, $-0.100 \pm 0.019 \text{ dex kpc}^{-1}$. 
All gradients are recorded in Table \ref{tab:azgradients} with the selection criteria, whether it traces a radial or azimuthal gradient, and how many clusters are included in the selection. 

\subsubsection{Azimuthal Gradients in Radius Slices}

\begin{figure}[t!]
    \begin{center}
    \epsscale{1.2}
 	\plotone{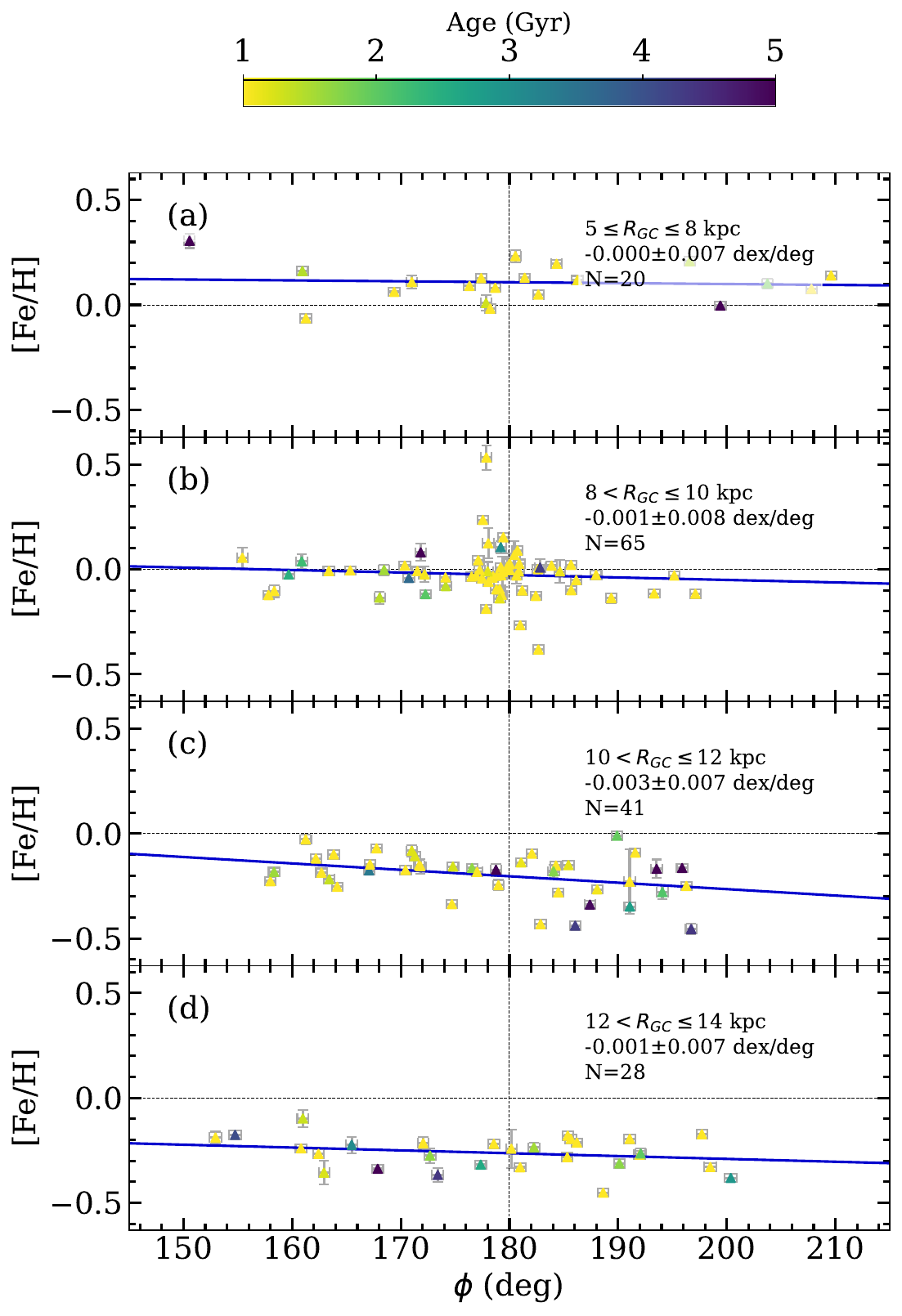}
 	\vskip0.1in
 	\caption{ \small Metallicity gradients [Fe/H] as a function of azimuth angle ($\phi$) for four slices in $R_{GC}$: (a) $5 \text{ kpc} \leq R_{GC} \leq 8$ kpc, (b) $8 \text{ kpc} < R_{GC} \leq 10$ kpc, (c) $10 \text{ kpc} < R_{GC} \leq 12$ kpc, and (d) $12 \text{ kpc} < R_{GC} \leq 14$ kpc. The linear fits (blue solid line), derived gradients, and the number of clusters (N) used are shown in each panel. Points are colored by the cluster age in Gyr, saturating at younger than 1 Gyr and older than 5 Gyr. Fits were determined using only clusters within the azimuth angle range $150\degree \leq \phi \leq 210\degree$. The regions corresponding to these radial slices are shown in Figure \ref{fig:XYaz}.}
 	\label{fig:az_azimuth}
 	\end{center}
\end{figure}

Figure \ref{fig:az_azimuth} shows the azimuthal [Fe/H] gradient determined for four slices (5--8 kpc, 8--10 kpc, 10--12 kpc, 12--14 kpc) in \rgc space. All four show extremely shallow, negative gradients that are consistent with flat trends, with the steepest being only $-0.003 \pm 0.007 \text{ dex deg}^{-1}$. The median [Fe/H] value for the clusters in each radius slice steadily decreases from a value of [Fe/H] $= 0.106$ dex in the 5--8 kpc slice to a median value of [Fe/H] $= -0.253$ in the 12--14 kpc radius slice. All azimuthal gradients and how many clusters were used in determining them are recorded in Table \ref{tab:azgradients}. 

\begin{deluxetable}{lrr}[t!]
\tabletypesize{\small}
\tablecaption{Radial and Azimuthal {[Fe/H]} Gradients \label{tab:azgradients}}
	\tablehead{
    \colhead{Selection} &
    \colhead{Gradient} &
    \colhead{ N} \\[-1ex]
    \colhead{} &
    \colhead{(dex kpc$^{-1}$)} &
    \colhead{} \\[-3.5ex]
    }
	\startdata
\multicolumn{3}{c}{d[Fe/H]/d$R_{GC}$}\\\hline
$210 \geq \phi \geq 195$  & $-0.057 \pm 0.022$ &  23 \\
$195 > \phi \geq 185$     & $-0.055 \pm 0.016$ &  28 \\
$185 > \phi \geq 175$     & $-0.093 \pm 0.009$ & 64 \\
$175 > \phi \geq 165$     & $-0.070 \pm 0.016$ & 26 \\ 
$165 > \phi \geq 150$     & $-0.076 \pm 0.026$ & 33 \\ \hline 
 \multicolumn{3}{c}{{d[Fe/H]/d$\phi$}}\\\hline
$5 \leq R_{GC} \leq 8 $  & $-0.000 \pm 0.007$ & 20 \\
$8 < R_{GC} \leq 10 $    & $-0.001 \pm 0.008$ & 65 \\
$10 < R_{GC} \leq 12$    & $-0.003 \pm 0.007$ & 41 \\
$12 < R_{GC} \leq 14$    & $-0.001 \pm 0.007$ & 28 
\enddata
\end{deluxetable}




\subsection{The Evolution of Galactic Abundance Gradients}
\subsubsection{Iron}\label{sec:fe_evo}

\begin{figure}[ht!]
    \epsscale{1.2}
 	\plotone{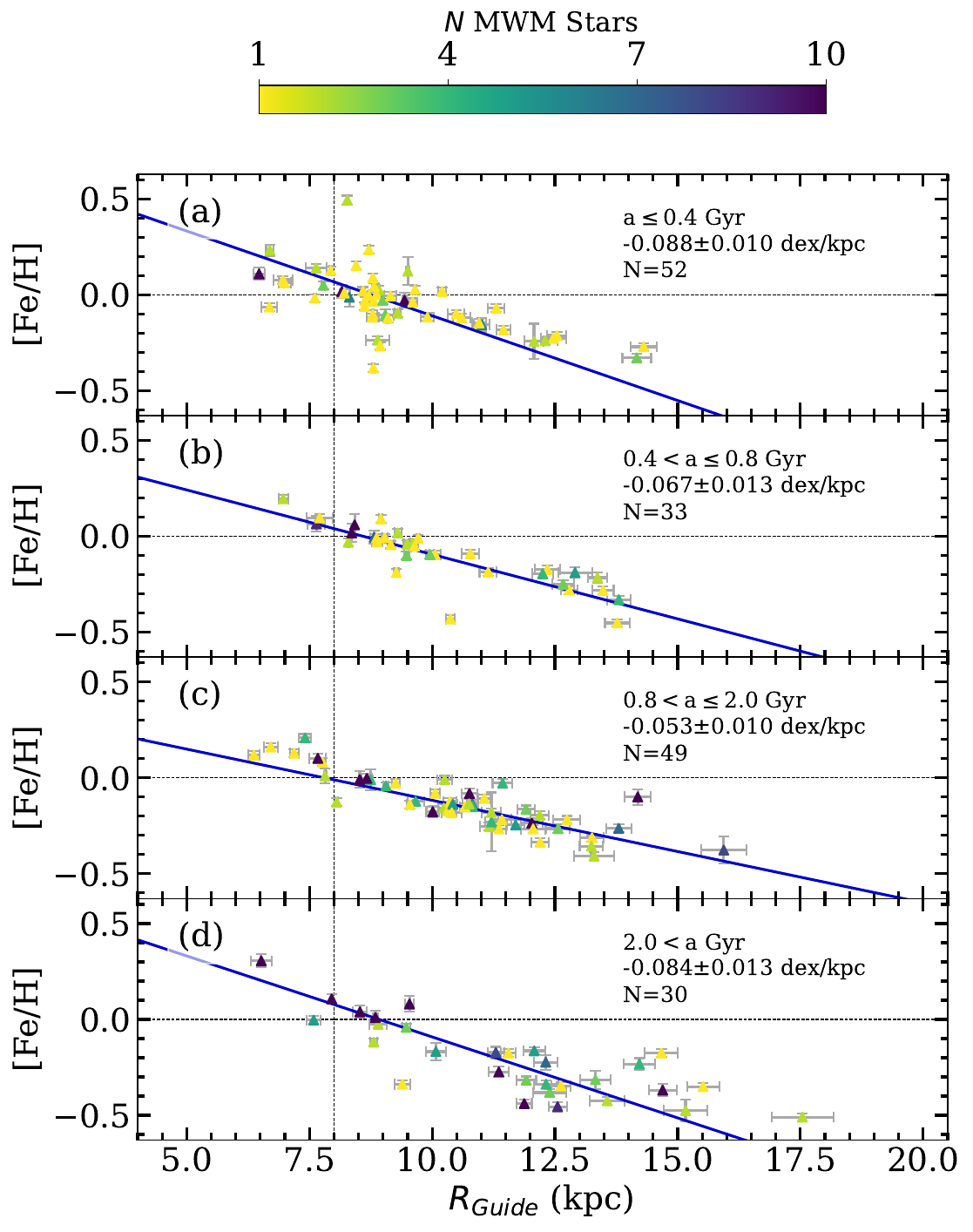}
 	\caption{ \small The Galactic [Fe/H] versus radius trend in four age bins: (a) age $\leq 0.4$ Gyr, (b) $0.4 \leq \text{age} \leq 0.8$ Gyr, (c) $0.8 \leq \text{age} \leq 2.0$ Gyr, and (d) age $> 2.0$ Gyr. Points colored by the number of stars in each cluster saturating at 10 stars. The derived linear gradient (solid blue line) and number of clusters (N) used for the fit are shown in each panel.}
 	\label{fig:gradage}
\end{figure}

The question of how Galactic metallicity gradients have evolved over the lifespan of the Milky Way is a prominent question that chemical evolution models are attempting to answer. 
As the OCCAM open cluster sample continues to grow, we become increasingly well equipped to address this question. 
We split our sample into 4 age bins, cluster ages derived in \citet{cg20}, younger than 400 Myr, 400-800 Myr, 800 Myr to 2 Gyr and older than 2 Gyr\footnote{These are the same bins as previous OCCAM analyses \citep{occam_p4,myers_2022} and the \citet{netopil22} study}. In Figure \ref{fig:gradage} we plot the [Fe/H] gradients with respect to \rguide in each of the four age bins.

All four age bins show clear negative gradients and three of the four gradients are within the uncertainty of the overall linear slope with respect to \rguide of $-0.074 \pm 0.010 \text{ dex kpc}^{-1}$. The youngest and oldest age bins have nearly identical slopes at $-0.088 \pm 0.010 \text{ dex kpc}^{-1}$ and $-0.084 \pm 0.013 \text{ dex kpc}^{-1}$ respectively. The second oldest age bin has a considerably shallower gradient at only $-0.053 \pm 0.010 \text{ dex kpc}^{-1}$. The second youngest age bin is between these two extremes at $-0.067 \pm 0.013 \text{ dex kpc}^{-1}$ and most closely matches the overall linear trend. 

The gradients for \rgc also all show clear negative trends, however, only two age bins agree with the overall linear slope with respect to \rgc, the second youngest and the oldest. The second oldest age bin has a considerably shallower gradient than the other three, matching the \rguide results. The youngest age bin has a considerably steeper gradient than the rest and more closely matches the inner slope of the two fit gradient. All [Fe/H] gradients and uncertainties for both radii in each age bin are shown in Table \ref{tab:fehgradients} along with how many clusters are used to determine the gradient.

\subsubsection{[X/Fe]}\label{elem_evo}

\begin{figure*}[t]
    \epsscale{1.1}
 	\plotone{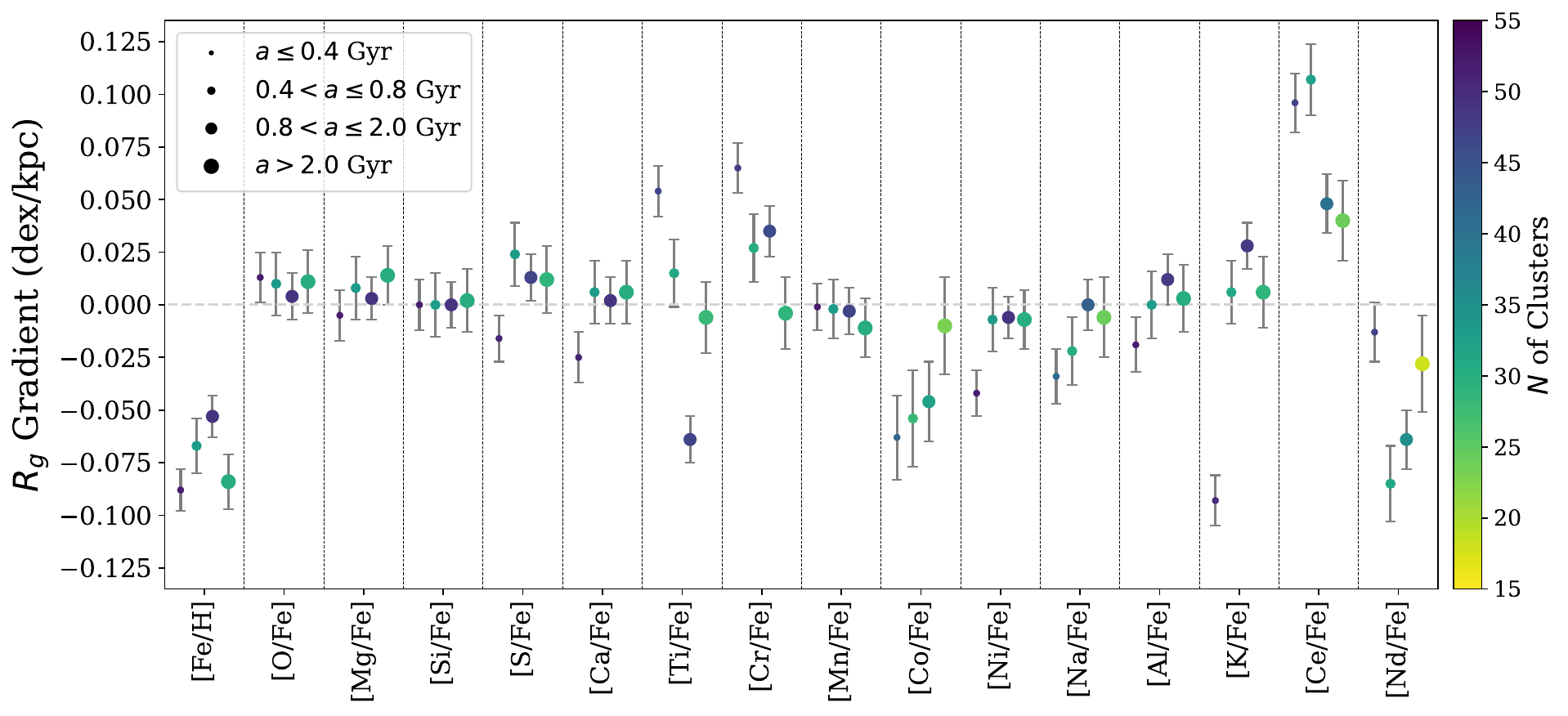}
 	\caption{ \small The slopes of each elemental gradient (d[X/Fe]/d\rguide or d[Fe/H]/d$R_{Guide}$) in four age bins (age bins defined as in Figure \ref{fig:gradage}). Point size increases with age, as indicated in the legend. Color indicates the number of clusters included in the gradient measurement, as shown by the color bar. }
 	\label{fig:slopesummary}
 \end{figure*}

In order to further investigate how radial chemical gradients have evolved over the lifetime of the Milky Way, we determine and report the gradients in the same four age bins as in section \ref{sec:fe_evo} for the 15 other elements included in this analysis. All gradients determined with respect to both \rguide and \rgc are reported in Table \ref{tab:gradients}. The gradients for each of the abundance ratios in all four age bins are shown in Figure \ref{fig:slopesummary}.    

\section{Discussion} \label{sec:discussion}


\subsection{Comparison to Other Surveys \label{sec:othercomp}}

In this section, we compare our open cluster sample to that of the previous OCCAM work \citep[][OCCAM-IV]{myers_2022} and three other large-scale high resolution spectroscopic surveys, 
the \gaia-ESO survey \citep[][]{gaia_eso}, the OCCASO survey \citep[][]{occaso}, and the GALAH survey \citep[][]{galah}. For each, a figure similar to Figure \ref{fig:compFe} was made, and the median offset and error were calculated with the set of clusters that were in common with our sample. 
%
\natp~released their catalog of 150 total open clusters, 94 of which were designated to be ``high-quality.'' A detailed comparison to the \natp~sample is provide in Appendix \ref{sec:dr17comp}. 
The \gaia-ESO survey released their catalog of 80 total open clusters, 62 science clusters and an additional 18 archive clusters. Of these 80 open clusters, 27 of them also appear in our sample, with a median offset of $-0.028$ dex with a scatter of $0.063$ dex. The latest update of the OCCASO survey \citep[][]{occasoV} comprises a sample of 36 open clusters, 15 of which are also in our sample. We show the best agreement with this set of clusters, with a median offset of only $+0.004$ dex and a scatter of $0.072$ dex. 
\citet{spina_21} curated a sample of 205 open clusters that have data for either GALAH or SDSS-IV/APOGEE, 94 of which can be found in our sample. We find a median offset of $-0.010$ dex with a measured scatter of $0.049$ dex. 
All of these offsets are well within the measured scatter of the respective comparison samples, indicating good general agreement between the different surveys for mean [Fe/H] values.  

\subsection{Comparison of Galactic Abundance Trends}

Comparisons against previous studies is an important step in evaluating the gradients we calculate and report in this work (see Tables \ref{tab:fehgradients}, \ref{tab:gradients}). We compare the gradient determined using the full sample of clusters with respect to \rgc because not all studies use $R_{Guide}$, and there is little to no difference in the gradient derived using the full sample versus only using the sample within \rgc of 14 kpc. 
While there is some agreement between the gradients calculated in this work and those of other studies, there is still no clear consensus. 
This suggests to the authors that cluster membership, sample size and composition, and abundance determination methodology still significantly influence the recovered overall Galactic trend.  

	

\subsubsection{Galactic Metallicity Gradient}

In this work we compute four total [Fe/H] radial gradients, a linear fit with respect to \rgc ($-0.075 \pm 0.006 \text{ dex kpc}^{-1}$) and \rguide ($-0.068 \pm 0.005 \text{ dex kpc}^{-1}$), as well as a two-component linear fit for \rgc with an inner slope of $-0.100 \pm 0.019 \text{ dex kpc}^{-1}$, a knee at $10.0 \pm 1.7$ kpc and an outer slope of $-0.044 \pm 0.036 \text{ dex kpc}^{-1}$. 
The two-component fit for \rguide was determined to have an inner slope of $-0.072 \pm 0.020 \text{ dex kpc}^{-1}$, a knee at $12.0 \pm 2.7$ kpc and an outer gradient of $-0.015 \pm 0.085 \text{ dex kpc}^{-1}$. 
Between this work and \natp, we find good agreement for the inner slope and knee using \rguide, though the outer slope is considerably shallower in this study. When looking at the gradients with respect to \rgc we find a steeper slope in both the inner and outer gradient than \natp~and the knee location has moved in by over 2 kpc. 

With the rise of large spectroscopic surveys, there have been numerous studies in recent years that characterize the radial metallicity gradient \citep[e.g.,][]{occasoV,magrini_23,gaia_grads,myers_2022,spina_21}. Literature gradient values range from $-0.048$ dex to $-0.076$ dex for a single linear fit using \rgc. The literature values along with the gradient from this work are shown in Table \ref{tab:litgrads} for linear gradients with respect to \rgc and inner and outer gradients where available. We note that this work found the linear [Fe/H] gradient to be steeper than the literature values, except for \citet{spina_21}, who found a gradient of $-0.076 \pm 0.009 \text{ dex kpc}^{-1}$ using 134 open clusters.
Samples that recover [Fe/H] gradient measurements similar to ours coincide with samples that have the most cluster overlap. 
This suggests to the authors that sample selection likely has a significant effect on gradient measurements and could explain the discrepancies seen between surveys. 

\begin{deluxetable*}{lrrrrcr}[t]
\tabletypesize{\small}
\tablecaption{Literature {[Fe/H]} Gradients \label{tab:litgrads}}
	\tablehead{
    \colhead{Reference} &
    \colhead{Linear Gradient} &
    \colhead{Inner Gradient} &
    \colhead{Outer Gradient} &
    \colhead{Knee} &
    \colhead{\rgc Range} &
    \colhead{ N} \\[-1ex]
    \colhead{} &
    \colhead{(dex kpc$^{-1}$)} &
    \colhead{(dex kpc$^{-1}$)} &
    \colhead{(dex kpc$^{-1}$)} &
    \colhead{(kpc)} &
    \colhead{(kpc)} &
    \colhead{} \\[-3.5ex]
    }
\startdata
\multicolumn{7}{c}{d[Fe/H]/d$R_{GC}$}\\\hline
This work           & $-0.075 \pm 0.006$  & $-0.100 \pm 0.019$ & $-0.044 \pm 0.036$ &  $10.0 \pm 1.7$  & 6--21  & 164 \\
OCCAM-IV \tablenotemark{a}              & $-0.055 \pm 0.001$  & $-0.073 \pm 0.002$ & $-0.032 \pm 0.002$ & $11.5 \pm 0.09$ & 6--18  & 85 \\
OCCASO+ \tablenotemark{b}    & $-0.062 \pm 0.007$  & $-0.069 \pm 0.008$ & $-0.025 \pm 0.011$ & $11.3 \pm 0.8\phn $ & 6--21  & 99 \\
\gaia-ESO \tablenotemark{c}  & $-0.054 \pm 0.004$  & $-0.081 \pm 0.008$ & $-0.044 \pm 0.014$ & \multicolumn{1}{l}{11.2} & 6--21 & 62 \\
ESA \gaia \tablenotemark{d}  & $-0.054 \pm 0.008$  & \multicolumn{1}{c}{\nodata}  & \multicolumn{1}{c}{\nodata}   & \multicolumn{1}{c}{\nodata} & 5--12 & 503 \\
GALAH DR3 \tablenotemark{e}   & $-0.076 \pm 0.009$  & \multicolumn{1}{c}{\nodata}   & \multicolumn{1}{c}{\nodata}   & \multicolumn{1}{c}{\nodata} & 6--17 & 134 \\
\enddata
\tablenotetext{a}{\citet{myers_2022}}\vskip-0.07in
\tablenotetext{b}{\citet{occasoV}}\vskip-0.07in
\tablenotetext{c}{\citet{magrini_23}}\vskip-0.07in
\tablenotetext{d}{\citet{gaia_grads}}\vskip-0.07in
\tablenotetext{e}{\citet{spina_21}}\vskip-0.07in
\end{deluxetable*}

\subsubsection{$\alpha-$Elements -- O, Mg, Si, S, Ca, Ti}\label{alphadiss}

In the [$\alpha$/Fe]-\rgc space, the gradients determined in this study agree with \natp~for three of the six elements studied in both works (Mg, Si, and S). We found a flat slope for [O/Fe] where \natp~had a shallow positive slope, and we find a shallow negative slope for [Ca/Fe] while \natp~had a shallow positive slope, though we note that our gradient of $-0.006 \pm 0.006 \text{ dex kpc}^{-1}$ is flat within $1\sigma$. The most stark difference is in [Ti/Fe] where \natp~report a shallow positive slope of $+0.004 \pm 0.002 \text{ dex kpc}^{-1}$, while here we calculated a negative gradient of $-0.019 \pm 0.006 \text{ dex kpc}^{-1}$. We agree well with the [Mg/Fe] and [Si/Fe] gradients reported in \citet{occasoV} using the OCCASO only sample, but differ considerably from the reported gradients for [Ca/Fe] and [Ti/Fe]. 

\citet{magrini_23} report gradients for five $\alpha$-elements, (O, Mg, Si, Ca, Ti). The gradients for [Mg/Fe] and [Si/Fe] agree well with the gradients in this study. 
They report a significantly steeper positive slope for [O/Fe] ($+0.048 \pm 0.009 \text{ dex kpc}^{-1}$) than found in both this work and \natp. The \citet{magrini_23} gradient values for Ca and Ti also differ considerably from this work. \citet{magrini_23} reports a shallow positive gradient for both [Ca/Fe], $+0.018 \pm 0.003 \text{ dex kpc}^{-1}$, and [Ti/Fe], $+0.012 \pm 0.003 \text{ dex kpc}^{-1}$, which are more in line with \natp~than this work, but still steeper than the relatively flat trends determined in \natp. 

\subsubsection{Iron-Peak Elements -- Cr, Mn, Co, Ni}\label{ironpeakdiss}

We differ considerably from \natp~in the iron-peak elements (Cr, Mn, Co, Ni) studied in both works. Extremely shallow, negative trends with respect to \rgc were found for all four iron-peak elements in \citet{myers_2022}, while we find that for Mn only. We find a positive trend in [Cr/Fe], ($+0.016 \pm 0.006 \text{ dex kpc}^{-1}$), which closely matches the result from \citet{occasoV}, ($+0.017 \pm 0.008 \text{ dex kpc}^{-1}$) and \citet{magrini_23}, ($+0.018 \pm 0.003 \text{ dex kpc}^{-1}$). We report negative gradients steeper than those in \natp~for both [Co/Fe] and [Ni/Fe]. Our gradients for [Co/Fe] and [Ni/Fe] are also considerably steeper than those reported in \citet{occasoV} and, \citet{magrini_23}, who each report essentially flat gradients. 

\subsubsection{Odd-Z Elements -- Na, Al, K}

There is good agreement between this work and \natp~for all three of the odd-z elements (Na, Al, K) that are in both studies. Our slopes for [Na/Fe] ($-0.027 \pm 0.007 \text{ dex kpc}^{-1}$), [Al/Fe] ($+0.008 \pm 0.007 \text{ dex kpc}^{-1}$), and [K/Fe] ($+0.011 \pm 0.006 \text{ dex kpc}^{-1}$) are well within the uncertainties of the slopes reported in \natp, $-0.021 \pm 0.006 \text{ dex kpc}^{-1}$ for [Na/Fe], $+0.009 \pm 0.002 \text{ dex kpc}^{-1}$ for [Al/Fe] and $+0.017 \pm 0.003 \text{ dex kpc}^{-1}$ for [K/Fe]. \citet{occasoV} reports gradients for two odd-z elements, Na and Al. They report a nearly identical slope for [Na/Fe] of $-0.027 \pm 0.008 \text{ dex kpc}^{-1}$ but a shallow, negative slope of $-0.013 \pm 0.007 \text{ dex kpc}^{-1}$ for [Al/Fe] in contrast to our shallow, positive slope. The gradient from \citet{magrini_23} for [Na/Fe] differs considerably from both this work and \citet{occasoV}, they report a nearly flat gradient of $+0.003 \pm 0.002 \text{ dex kpc}^{-1}$. The gradient for [Al/Fe], $+0.012 \pm 0.004$ agrees well with our value.  

\subsubsection{Neutron Capture Elements -- Ce, Nd}\label{neutrondiss}

Cerium was the only neutron capture element measured and reported in DR17, and both this work and \natp~report positive slopes, though this work finds a considerably steeper value of $+0.087 \pm 0.007 \text{ dex kpc}^{-1}$ than \natp, $+0.022 \pm 0.006 \text{ dex kpc}^{-1}$. The steeper value found in this work is also in contrast with the reported values from \citet{occasoV}, $+0.001 \pm 0.013 \text{ dex kpc}^{-1}$ and \citet{magrini_23}, $+0.014 \pm 0.003 \text{ dex kpc}^{-1}$. We find a negative gradient for [Nd/Fe] of $-0.052 \pm 0.008 \text{ dex kpc}^{-1}$ that is starkly different from the positive slope gradients reported by \citet{occasoV}, $+0.032 \pm 0.016 \text{ dex kpc}^{-1}$ and \citet{magrini_23}, $+0.045 \pm 0.006 \text{ dex kpc}^{-1}$.  

\subsection{Azimuthal Gradients}\label{azgradsdiss}

We leveraged our robust sample of open clusters to investigate whether the radial metallicity gradient varies with azimuth angle. We constructed a grid in the X-Y plane of the Galaxy, encompassing a range of Galactocentric radii, ($5 \text{ kpc} \leq R_{GC} \leq 14 \text{ kpc}$, and azimuth angles, $150\degree \leq \phi \leq 210\degree$, as shown in Figure \ref{fig:XYaz}. The radial gradients in the azimuth slices and the azimuthal gradients in the radius slices are reported in Section \ref{azgrads} and summarized in Table \ref{tab:azgradients}. These results allow us to tentatively suggest that there might be real variations in the radial gradient as the azimuth angle changes. The radial gradients in the two azimuth bins with angles larger than $180 \degree$ (the solar location) have shallower gradients than the two bins with angles less than $180 \degree$; all four are shallower than the center bin where the Sun resides. We note the large uncertainty on all of these slopes, barring the Solar bin, due to the small number of clusters in each bin. Increasing the number of clusters available for this analysis will help clarify these trends. 

When calculating the azimuthal gradient in slices of constant radius, we see flat trends in all four slices. This lack of trend in azimuth angle was also seen in open clusters by \citet{occasoV} in both the OCCASO-only sample and the OCCASO+ sample. This result closely matches the results from \citet{hackshaw_2024}, who investigated azimuthal variations in the [Fe/H]-\rgc space using field stars from APOGEE/DR17. They found deviations from the overall [Fe/H] radial gradient across the disk, but lines of fairly consistent deviation that follow lines of constant radii. 
They note that while some studies have suggested the deviations track spiral arm structure \citep[][]{hawkins_23,poggio_22}, they do not find such a strong correlation. 
Using a large sample of young clusters will bring more context to this difference, which we aim to do in an upcoming study (Otto et al. 2025, {\it in prep}) using the MWM/BOSS data.  

\subsection{Evolution of Galactic Abundance Gradients}

\subsubsection{Iron}\label{sec:fe_evo_diss}

With the increase in the number of clusters in our sample which cover a wide range of ages, ($\sim 7$ Myr -- $\sim 7$ Gyr), we are well positioned to quantify how the radial [Fe/H] gradient has evolved over time, which is a core focus of the \occam survey. While we do see some variation in the [Fe/H] gradient, it is not as conclusive as previous studies have shown, \citep[e.g.,][]{myers_2022,spina_21, Netopil21, Zhang21}. When determining the gradient with respect to \rguide, we find the oldest and youngest age bins have nearly identical slopes ($-0.088 \pm 0.010 \text{ dex kpc}^{-1}$ and $-0.084 \pm 0.013 \text{ dex kpc}^{-1}$, respectively). All age bins except the 0.8 Gyr--2 Gyr bin are within the uncertainty of the overall linear slope of $-0.074 \pm 0.005 \text{ dex kpc}^{-1}$. Similar results are seen when \rgc is used as the independent variable. The youngest and oldest bins are the steepest, while the second-oldest bin is the shallowest. 

To compare our [Fe/H]-\rgc gradient evolution results with those of \citet{magrini_23} and \citet{occasoV}, which both performed similar analyses, we binned our clusters using the age bins described in each. \citet{magrini_23} used three age bins, $a \leq 1$ Gyr, $1 < a \leq 3$ Gyr and $a > 3$ Gyr. Our results agree well with theirs for the middle and oldest age bins, but differ significantly for the youngest age bin. They find a definitive trend where the youngest age bin has the shallowest slope, which steepens as age increases. We find a similar result to our own, where the youngest and oldest age bins are very similar and both are steeper than the middle age bin. 
Once again, all three gradients are within the uncertainty of the overall linear trend with $R_{GC}$. \citet{occasoV} used four age bins $0.2 < a \leq 1$ Gyr, $1 < a \leq 2$ Gyr, $2 < a\leq 3$ Gyr and $3 < a \leq 7.3$ Gyr. We find good agreement with the slopes for the oldest and second-youngest age bins and moderate agreement for the second-oldest age bin, though we note that due to the small number of clusters in that age bin (11 for our sample), there is a $0.026 \text{ dex kpc}^{-1}$ uncertainty in the gradient derived for that bin. They report the steepest slope for the oldest age bin and the shallowest for the second-oldest age bin, similar to what we find in our age bins. Though we differ considerably on the youngest age bin, the gradient calculated in this study is considerably steeper than theirs, both with their age bin and ours. 

\subsubsection{[X/Fe]}\label{elem_evodiss}

We leveraged the full ensemble of elements available from MWM/DR19 to explore how radial gradients in other elements, in addition to Fe, have evolved. \natp~also performed a similar analysis and found no significant trends in the $\alpha$-elements. We find the same, except for a possible trend in [Ti/Fe], which appears to be anti-correlated with [Fe/H]. Although we note there is a large amount of uncertainty in Ti gradients due to the relatively large amount of scatter in the element, as discussed in Section \ref{alpha}. For the iron-peak elements, we see a possible trend in [Mn/Fe] and [Cr/Fe] where the gradient steepens as a function of age and a possible trend in [Co/Fe] where the gradient gets shallower as we move from the younger to the older age bins. However, we note the relatively large errors on both [Cr/Fe] and [Co/Fe]. \natp~found no convincing trends in the iron-peak elements. 


\natp~noted a possible trend in [Na/Fe] where the gradient became steeper as clusters got younger; that same possible trend is present here and also in another odd-z element, Al. We do not find a smooth trend in the neutron capture element Ce, but there is a significant difference in the gradients between the two youngest age bins and the two oldest age bins. 
There does appear to be a possible trend in [Nd/Fe] where the gradient steepens for the younger clusters, particularly when accounting for the youngest age bin, which appears to be an outlier for close to half of the elements studied, including Nd.

\subsection{Comparison to Galactic Chemical Evolution Models}

\subsubsection{Description of the Models}



\begin{figure*}[t!]
    \epsscale{1.22}
    \plotone{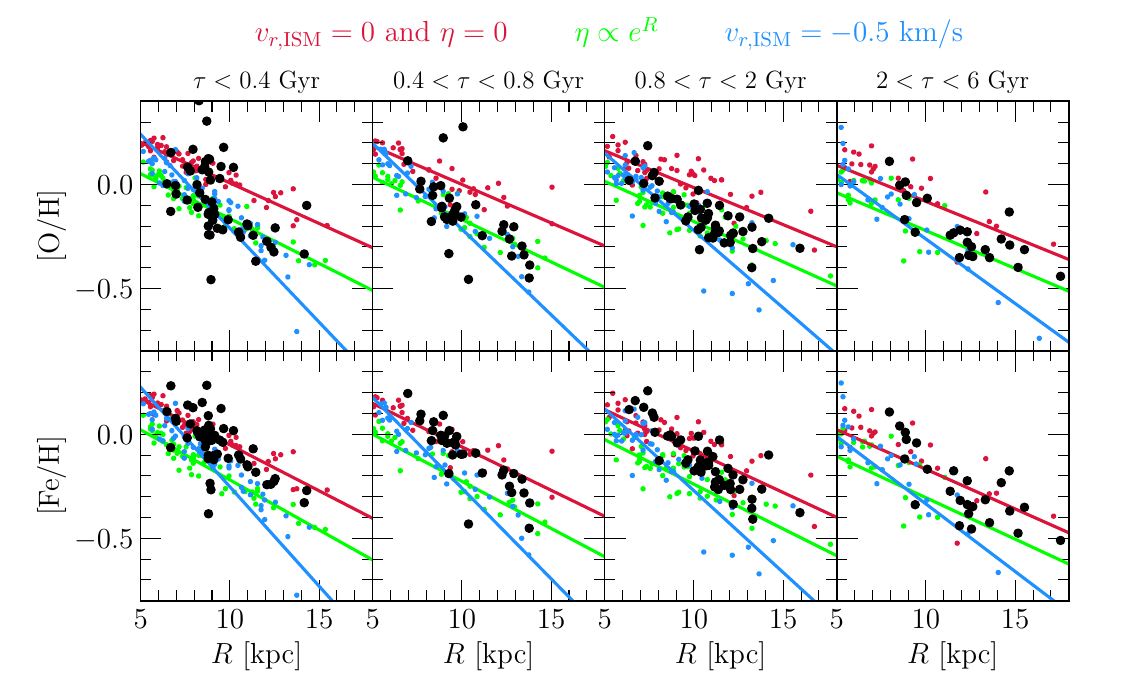}
        \caption{
        A comparison of the OCCAM cluster radial metallicity profiles with GCE models from \citet{Johnson2024} and Johnson et al. (2025, in prep.).
        The top panels show [O/H] versus Galactocentric radius (R), and bottom panels show [Fe/H] versus R. 
        The OCCAM data are shown as black points.
        Colored lines show the mass-weighted abundance profiles predicted by each GCE model: no outflows or radial gas flows (red), with outflows only (lime green), and with radial gas flows only (blue).
        Colored points show a random subsample of stars equal to the number of OCCAM clusters in each panel from the corresponding model.
        }
	\label{fig:gce-models}
\end{figure*}

In this section, we compare our results to a handful of GCE models that recently appeared in the literature.
\citet{Johnson2024} built on their previous work \citep{Johnson2021}, which discretized the disk into a series of 200 rings, each with width $\delta R = 100$ pc.
Each ring is coupled to its neighbors through radial migration, which exchanges stellar populations between them, but is otherwise described by a conventional one-zone GCE model (see, e.g., the reviews by \citealt{Tinsley1980} and \citealt{Matteucci2021}).
In \citet{Johnson2024}, the authors focused on reproducing recent results indicating that metallicity does not decline substantially with stellar population age, even for stars as old as $\sim$$8 - 9$ Gyr \citep[e.g.,][]{Spina2022, daSilva2023, Willett2023, Gallart2024}.
They argued that outflows ejecting ISM material to the CGM are one possible origin of this behavior.
We use their model with an exponential dependence on radius for the outflow mass loading factor, $\eta \equiv \dot \Sigma_\text{out} / \dot \Sigma_\star$, which describes the rate of mass ejection relative to star formation.
For comparison, we also use their model with $\eta = 0$ everywhere.
\par
J. W. Johnson et al. (2025, in preparation) explore an extension of these models in which the outflow is replaced with a radial gas flow.
These flows are generally thought to be directed inward, carrying gas toward the centers of disk galaxies \citep[see discussion in, e.g.,][]{Bilitewski2012}.
J. W. Johnson et al. (2025, in preparation) explore multiple assumptions about what processes drive the radial gas flow.
We also compare our measurements with their simplest prescription, which assumes a velocity in the ISM that is constant in both radius and time, taking $v_{r,\text{ISM}} = -0.5 \text{ km s}^{-1}$.

\subsubsection{OCCAM-GCE Model Comparison}

Figure \ref{fig:gce-models} compares the OCCAM cluster [O/H] and [Fe/H] radial profiles in different age bins with the predictions of these GCE models.
In [O/H], the outflow-driven model from \citet{Johnson2024} tentatively offers the best explanation for the OCCAM data across all age bins.
The model with neither outflows nor radial gas flows overpredicts [O/H] at all ages.
This difference is a natural consequence of outflows, which lower abundances by removing metal-rich material from the ISM and replacing it with metal-poor gas through accretion.
In [Fe/H], however, the OCCAM data tentatively favor the model with neither outflows nor radial flows for its higher normalization.
The radial gas flow model underpredicts [Fe/H] overall and [O/H] in the oldest age bins.
In qualitative agreement with previous work \citep[e.g.,][]{Spitoni2011}, the radial gas flow leads to a steep radial gradient, so the underprediction is most obvious in the outer disk.
\par

Model uncertainties and a small sample prevent us from definitively favoring any one model over another.
In particular, a slight increase in Fe yields could place the outflow-driven model from \citet{Johnson2024} at a slightly higher normalization, thereby improving its agreement with the OCCAM data.
A decrease in O yields would similarly lower the predicted [O/H] abundances overall, bringing the $\eta = 0$ and $v_{r,\text{ISM}} = 0$ model into better agreement.
Each of these models assumes the same overall normalization of stellar yields recommended by \citet{Weinberg2024}, which in turn is based on the analysis of the radioactive tails of Type II supernova light curves by \citet{Rodriguez2023}.
However, the statistical uncertainty of this recommendation is roughly $\sim$$0.1$ dex, which is comparable to the difference between models in Figure \ref{fig:gce-models}.
Vertical shifts of the radial gas flow model at this level would also improve agreement with the data.
\par
Previous reports from the OCCAM survey \citep{myers_2022, occam_p4} have compared their data with the thin disk GCE models from \citet{chiappini_2009} and \citet{minchev_1, minchev_2}.
These earlier models omit both outflows and radial gas flows, so they equate most directly to the model from \citet{Johnson2024} that also assumes $\eta = 0$ and $v_{r,\text{ISM}} = 0$.
The previous models predicted radial metallicity profiles consistent with the data in all age bins except the oldest ($2 - 6$ Gyr), in which they underpredicted metallicity overall by $\sim$$0.2$ dex (see Figure 13 of \citealt{occam_p4} and Figure 14 from \citealt{myers_2022}).
The models from \citet{Johnson2024} appear to resolve this issue, predicting slightly higher abundances in this age range in reasonable agreement with the OCCAM data.
The origin of this improvement upon previous models is unclear and outside the scope of this paper.



\section{Conclusions }\label{sec:conclusions}

We present a sample of 164 quality open clusters which comprises the full OCCAM MWM/DR19 sample. 
We leverage this large sample to investigate the radial Galactic gradient for 16 elements both overall and in four age bins. We do not find convincing trends with age for any of the elements. 
In this work, we compute a bilinear fit in addition to a linear fit for the [Fe/H] gradient both for \rgc and $R_{Guide}$.  To determine which model (linear or bilinear) best fits the data, we conduct an AIC analysis and find that the linear fit provides the best fit for both \rgc and $R_{Guide}$. 
We find a linear metallicity gradient of $-0.075 \pm 0.006 \text{ dex kpc}^{-1}$ with respect to $R_{GC}$. And a linear metallicity gradient of $-0.068 \pm 0.005 \text{ dex kpc}^{-1}$ with respect to $R_{Guide}$. 
Additionally, we find evidence for azimuthal variations across the disk in the open clusters, though the extent to which that is driven by spiral structure is unclear with the current sample. 

While we find good agreement with the average cluster abundances for the common clusters between our sample and the samples from \gaia-ESO, LAMOST, GALAH, and OCCASO, the gradients themselves differ considerably for some elements (e.g., O, Ca, Ti, Co, Ni). 
This descrepancy may be attributed to differences in abundance measurement between the surveys; however, we also suggest that it could larely be due to variations in sample composition mojng the various open cluster catalogs used in each respective work.  

When comparing to Galactic evolution models, we tentatively suggest that the [O/H] abundances prefer a model that includes gas outflows, while [Fe/H] seems to prefer a model with no gas outflows or radial gas flows.  More data and a more robust exploration of model fits is needed to truly constrain what model best fits the open cluster data.

\begin{acknowledgements}



We would like to thank Alexandre Roman Lopes and Chritian Moni Bidin for helpful comments. 

We acknowledge support for this research from the National Science Foundation Collaborative Grants (AST-2206541, AST-2206542, and AST-2206543) awarded to PMF, GZ, and KC respectively, that supported (JMO, PMF, NRM, AH, JD, PMF, AIW, TS, GZ, and KC).  PMF, SL, GZ, and APW acknowledge some of this work was performed at the Aspen Center for Physics, which is supported by National Science Foundation grant PHY-1607611. JWJ acknowledges support from a Carnegie Theoretical Astrophysics Center postdoctoral fellowship. SM has been supported by the LP2021-9 Lend\"ulet grant of the Hungarian Academy of Sciences and by the NKFIH excellence grant TKP2021-NKTA-64.\\

Funding for the Sloan Digital Sky 
Survey IV has been provided by the 
Alfred P. Sloan Foundation, the U.S. 
Department of Energy Office of 
Science, and the Participating 
Institutions. 

SDSS-IV acknowledges support and 
resources from the Center for High 
Performance Computing  at the 
University of Utah. The SDSS 
website is www.sdss.org.

SDSS-IV is managed by the 
Astrophysical Research Consortium 
for the Participating Institutions 
of the SDSS Collaboration including 
the Brazilian Participation Group, 
the Carnegie Institution for Science, 
Carnegie Mellon University, Center for 
Astrophysics | Harvard \& 
Smithsonian, the Chilean Participation 
Group, the French Participation Group, 
Instituto de Astrof\'isica de 
Canarias, The Johns Hopkins 
University, Kavli Institute for the 
Physics and Mathematics of the 
Universe (IPMU) / University of 
Tokyo, the Korean Participation Group, 
Lawrence Berkeley National Laboratory, 
Leibniz Institut f\"ur Astrophysik 
Potsdam (AIP),  Max-Planck-Institut 
f\"ur Astronomie (MPIA Heidelberg), 
Max-Planck-Institut f\"ur 
Astrophysik (MPA Garching), 
Max-Planck-Institut f\"ur 
Extraterrestrische Physik (MPE), 
National Astronomical Observatories of 
China, New Mexico State University, 
New York University, University of 
Notre Dame, Observat\'ario 
Nacional / MCTI, The Ohio State 
University, Pennsylvania State 
University, Shanghai 
Astronomical Observatory, United 
Kingdom Participation Group, 
Universidad Nacional Aut\'onoma 
de M\'exico, University of Arizona, 
University of Colorado Boulder, 
University of Oxford, University of 
Portsmouth, University of Utah, 
University of Virginia, University 
of Washington, University of 
Wisconsin, Vanderbilt University, 
and Yale University.\\

Funding for the Sloan Digital Sky Survey V has been provided by the Alfred P. Sloan Foundation, the Heising-Simons Foundation, the National Science Foundation, and the Participating Institutions. SDSS acknowledges support and resources from the Center for High-Performance Computing at the University of Utah. SDSS telescopes are located at Apache Point Observatory, funded by the Astrophysical Research Consortium and operated by New Mexico State University, and at Las Campanas Observatory, operated by the Carnegie Institution for Science. The SDSS web site is \url{www.sdss.org}.

SDSS is managed by the Astrophysical Research Consortium for the Participating Institutions of the SDSS Collaboration, including Caltech, The Carnegie Institution for Science, Chilean National Time Allocation Committee (CNTAC) ratified researchers, The Flatiron Institute, the Gotham Participation Group, Harvard University, Heidelberg University, The Johns Hopkins University, L'Ecole polytechnique f\'{e}d\'{e}rale de Lausanne (EPFL), Leibniz-Institut f\"{u}r Astrophysik Potsdam (AIP), Max-Planck-Institut f\"{u}r Astronomie (MPIA Heidelberg), Max-Planck-Institut f\"{u}r Extraterrestrische Physik (MPE), Nanjing University, National Astronomical Observatories of China (NAOC), New Mexico State University, The Ohio State University, Pennsylvania State University, Smithsonian Astrophysical Observatory, Space Telescope Science Institute (STScI), the Stellar Astrophysics Participation Group, Universidad Nacional Aut\'{o}noma de M\'{e}xico, University of Arizona, University of Colorado Boulder, University of Illinois at Urbana-Champaign, University of Toronto, University of Utah, University of Virginia, Yale University, and Yunnan University.\\

This work has made use of data from the European Space Agency (ESA) mission \gaia (\url{https://www.cosmos.esa.int/gaia}), processed by the \gaia Data Processing and Analysis Consortium (DPAC, \url{https://www.cosmos.esa.int/web/gaia/dpac/consortium}). Funding for the DPAC has been provided by national institutions, in particular the institutions participating in the \gaia Multilateral Agreement.

This research made use of Astropy, a community-developed core Python package for Astronomy (Astropy Collaboration, 2018).
\end{acknowledgements}

\begin{contribution}
JMO led the DR19 analysis and is the primary author of the paper.
PMF provided overall guidance of the OCCAM survey and contributed to the data analysis and writing of the paper.
NRM provided analysis and comparison to DR17 OCCAM survey.
KC provided expertise in evaluating the quality of the abundances.
JWJ provided comparison to the galaxy evolution modeling.
AH, HW, \& DS contributed to the data analysis and writing of the paper
JD contributed to the development of the OCCAM pipeline
AIW, TS, SL, \& GZ contributed to evaluation of the OCCAM sample.
AMPW contributed to the Galactic orbital computations and analysis.\\
SDSS-V Architects (Dmitry Bizyaev, Kaike Pan and Andrew Saydjari) contributed more than 1 FTE to the fund-raising, proposal writing, hardware, software, engineering, operations, observing, data archiving, and/or scientific organization of SDSS-V from which the data from this work are derived.
\end{contribution}

\facilities{Du Pont (APOGEE), Sloan (APOGEE), Spitzer, WISE, 2MASS, Gaia}
\software{\href{http://www.astropy.org/}{Astropy}, \href{https://emcee.readthedocs.io/en/stable/}{emcee}, \href{https://gala.adrian.pw/en/latest/index.html}{gala}}

\appendix 
\section{Detailed Comparison to \natp~DR17 OCCAM sample\label{sec:dr17comp}}

We recover 93 of the clusters in \natp~and add another 77 clusters. Due to methodology differences between this work and \natp, there are 60 clusters that we do not recover, including 16 of the clusters deemed ``high quality'' by \natp. The vast majority of these clusters had only a single member star. The reason these clusters are not included in the DR19 OCCAM run can be attributed to one of three things. First, methodological differences between this work and previous OCCAM papers (detailed in Section \ref{methoddiff}) are a factor. Specifically, using the \citet{cg_18} membership as a replacement for the proper motion analysis of previous OCCAM papers resulted in a much more restrictive membership probability cut ($>70$\%), which excludes a number of stars included in \natp. 
A thorough discussion of the differences between the previous OCCAM methodology and the \citet{cg_18} methodology can be found in \natp. Second, there are clusters that we believe to be real but are too distant from the Sun to be recovered by \citet{cg_18}, e.g., Saurer 1. The final reason that a cluster may no longer be included in our sample, particularly for the low quality sample from \natp, is that it is may not be a real cluster or the observed star should not be considered a member.

\begin{figure}[h!]
    \epsscale{1.15}
    \plotone{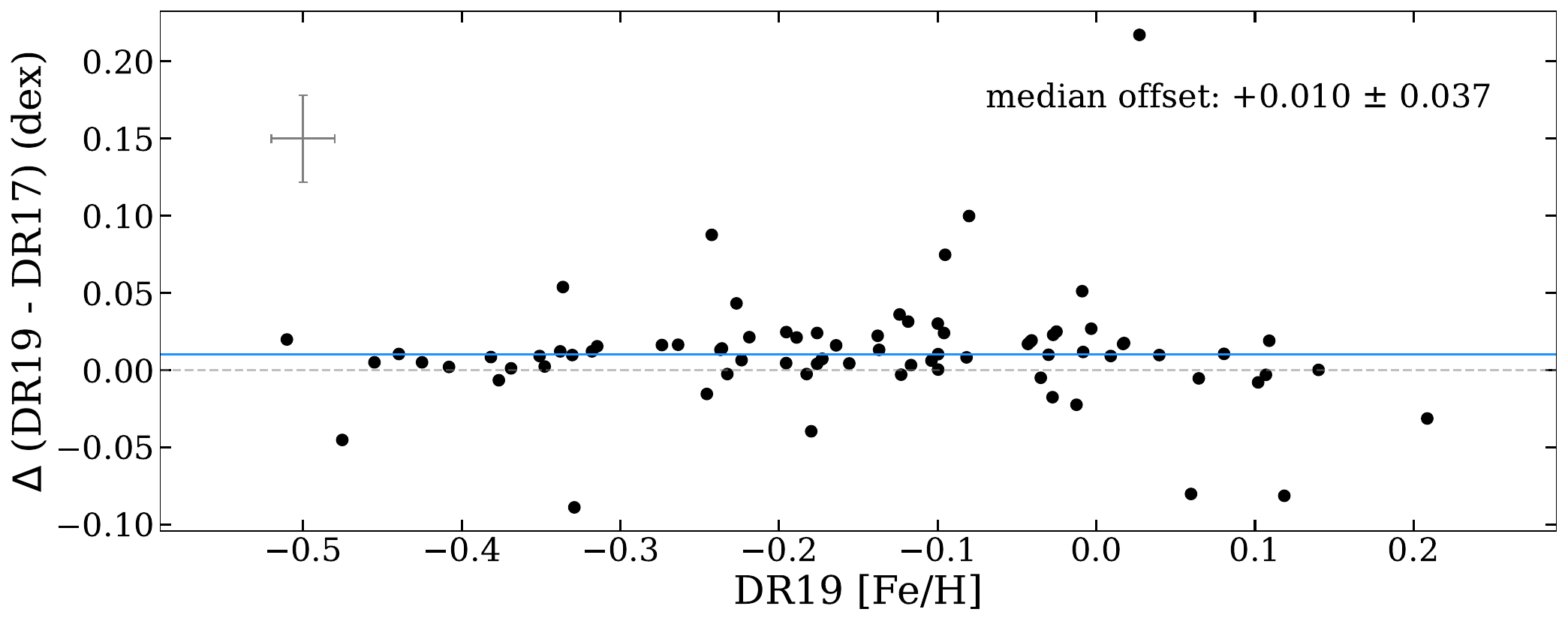}
 	\caption{ \small Comparing the MWM/DR19 (this work) and APOGEE/DR17 (\natp) bulk cluster [Fe/H] abundances. The measured median offset ($+0.010 \pm 0.037$ dex) is indicated by the solid blue line, while the grey dashed line shows the zero-difference point. A median characteristic error bar for the data points is shown in the top left corner of the figure.}
 	\label{fig:compFe}
\end{figure}

Of the 93 clusters recovered from \natp~in this work, a total of 79 were deemed to be high-quality in \natp. For the sample in common, we show the change in [Fe/H] from the DR17 ASPCAP abundance to the DR19 ASPCAP abundance in Figure \ref{fig:compFe}. The median offset between the two datasets is measured to be $+0.010$ with a $1\sigma$ scatter of 0.037, which is considerably larger than the median offset. Previous comparisons between the open cluster [Fe/H] abundances for consecutive data releases have shown that the scatter is primarily due to the metal-poor clusters, but here we note that significant scatter is seen over the full range of [Fe/H] abundances. 
Membership differences can explain this small offset and increased scatter in the more metal-rich clusters due to methodology chnages and changes in the ASPCAP pipeline, disussed in Casey et al. 2025 {\it in prep}.

\begin{figure*}[]
    \epsscale{0.95}
    \plotone{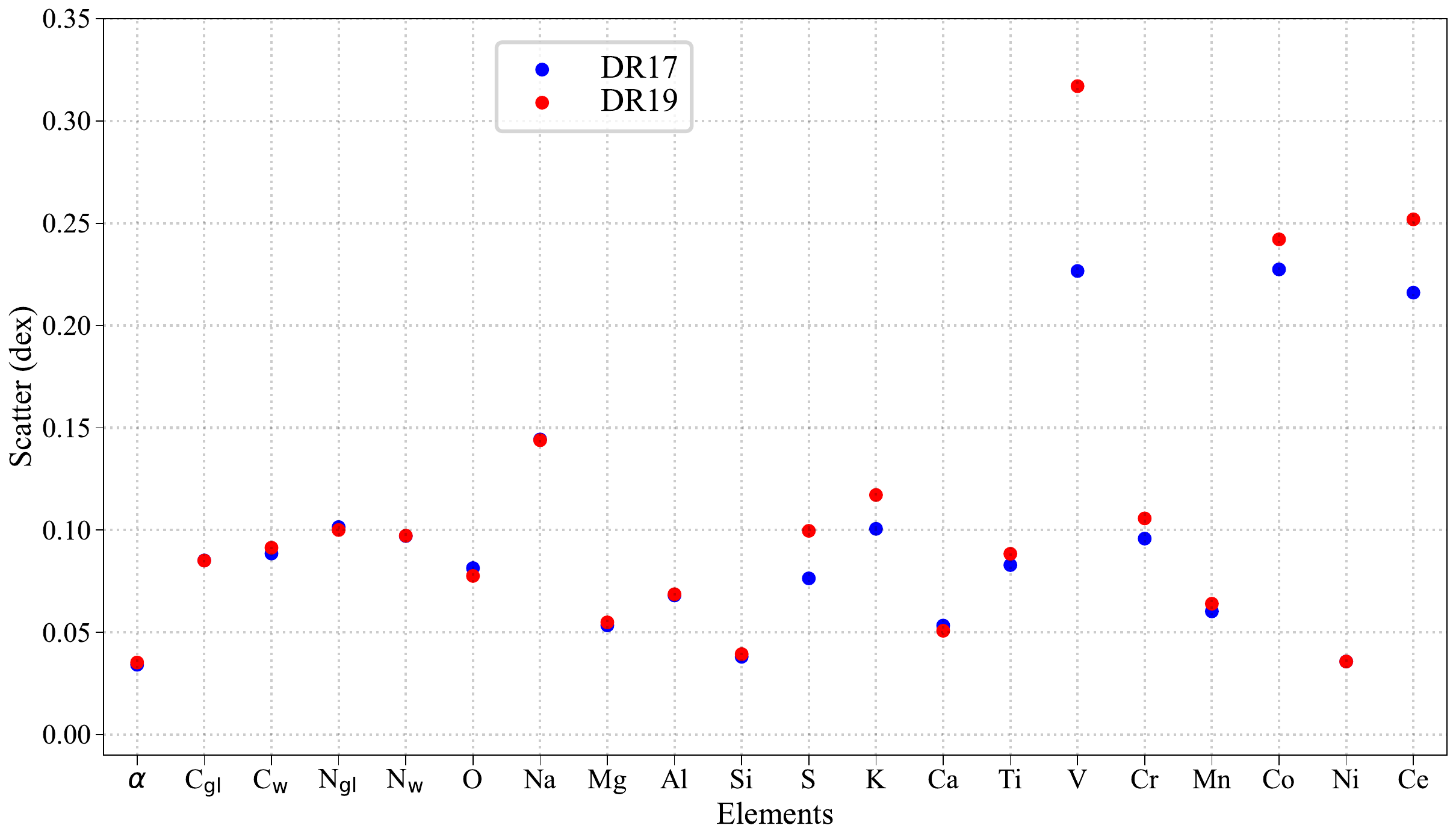}
	\caption{ \small Scatter of raw abundances using 526 member stars in common between the OCCAM sample from DR17 (blue dots) with the DR19 ones (red dots). $N_{\rm gl}$ and $C_{\rm gl}$ are [N/M] and [C/M] values derived from the global fit of spectra,$N_{\rm w}$ and $C_{\rm w}$ are derived from spectral windows centered around CN and CO lines, respectively. While the scatter for most elements are very similar, slightly elevated scatter can be seen in case of S, K, Tu, V, Cr, Co, and Ce. }
	
	\label{fig:starbystar}
\end{figure*}

When using the {\em same} cluster member stars, the scatter of abundances can be used to assess the overall precision. Here, we compare the precision of raw abundances from 18 species published both in DR17 and DR19 in Figure \ref{fig:starbystar} in a sample consisting of 526 stars considered to be cluster members with high confidence. The precision of multiple elements can be considered the same in DR19 and DR17, including C, N, O, Na, Mg, Al, Si, Ca, and Ni. It appears, however, that abundances of S, K, Tu, V, Cr, Co, and Ce have slightly worse precision in DR19 than DR17 to a varying degree. Among these, V and Ce seem to be affected the most. The scatter of V increased from 0.227 to 0.317~dex, while Ce increased from 0.216 to 0.252~dex. 
The change in precision for the other elements is minor. 
The common property of the affected elements is that all of these species have weak lines in the $H$-band, making it generally difficult to measure their abundances with high precision. While we do not know the exact cause behind the slightly decreased precision, it could be that some subtle change in the way the spectra is processed in DR19 causes this issue. A detailed analysis of the accuracy and precision of abundances published in DR19 can be found in S. Mészáros et al. 2025 \url{(https://ui.adsabs.harvard.edu/abs/2025arXiv250607845M/abstract)}.

\begin{figure*}[b]
    \epsscale{0.95}
    \plotone{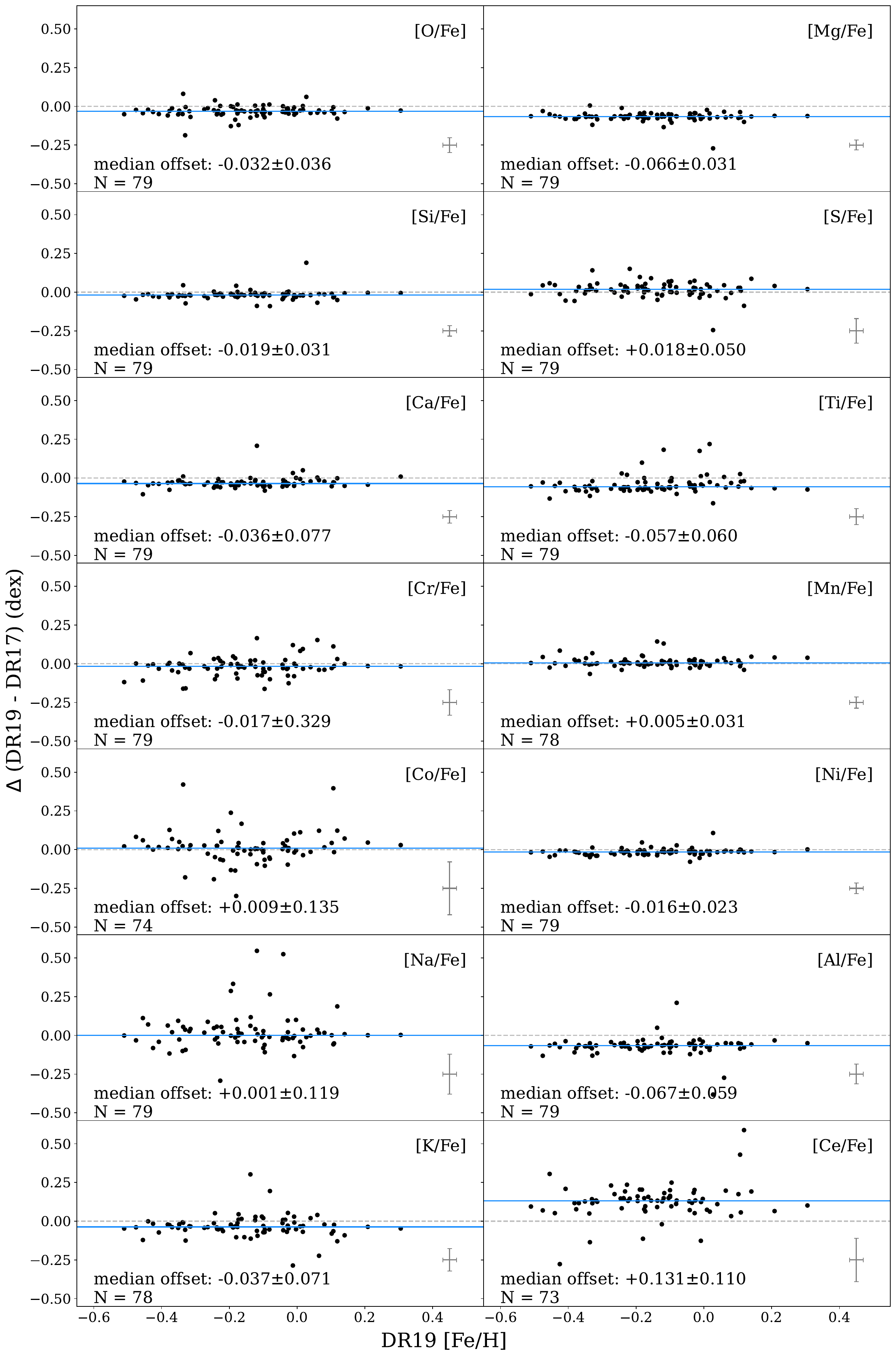}
	\caption{ \small Comparison of bulk cluster abundance ratios ([X/Fe]) between MWM/DR19 (this work) and APOGE/DR17 \citep{myers_2022} comparison for 14 elements, plotted against the DR19 [Fe/H] abundance of the clusters. The median offset is indicated by the solid blue line, while the grey dashed line shows the zero point. A median characteristic error bar is shown in the bottom right of each subplot. }
	
	\label{fig:dr19v17ab}
\end{figure*}

Figure \ref{fig:dr19v17ab} shows the DR19 vs DR17 bulk cluster abundances for 14 of the remaining elements in this study; neodymium is left out since it was not reported in DR17. 11 of the 14 elements analyzed in this study have median offsets within the measured scatter for that element that can be explained by membership differences and changes in the ASPCAP pipeline from DR19 to DR17. Of the three that do not (Mg, Ni, and Ce), Ni has a small offset with a slightly higher scatter that is heavily influenced by a few outliers not shown on the plot. Ce has a large offset and a large scatter, and is considered a less reliable element in DR19. The last one, Mg, has an offset of $-0.066$ dex with a scatter of $0.031$ dex, suggesting there is a real systematic offset between the DR19 and DR17 Mg abundances, likely due to the inclusion of NLTE corrections in DR17 \citep{dr17} that were not included for DR19. 

\clearpage
\section{Open Cluster Sample with \citet{HUNT_2023} Membership}\label{ehmems}

Here we provide the open cluster sample using the \citet{HUNT_2023} catalog (EH catalog) as the starting point of the analysis. 
Following the same procedure laid out in Section \ref{sec:methods}, we start by selecting all stars in MWM/DR19 that are within $R_{tot}$\footnote{The total radius of the cluster, including tidal tails and the coma as defined by \citet{HUNT_2023}.} of the cluster center.  Using this subset, we select our proper motion member stars by keeping only the stars that have a >5\% probability of being a cluster member. 
This subset of stars is the proper motion member stars that we start our analysis with. 
We then compute the RV and metallicity membership probabilities for each of the proper motion members using the MWM/DR19 RVs and [Fe/H] abundances. 
We apply a Gaussian kernel smoothing routine in both parameters spaces (RV and [Fe/H]). 
A Gaussian distribution is then fit to the distribution and normalized to compute the membership probabilities.
Stars with a membership probability above 5\% in all three parameter spaces are used to calculate bulk cluster parameters. 

\begin{figure}
  \epsscale{1}
 \plotone{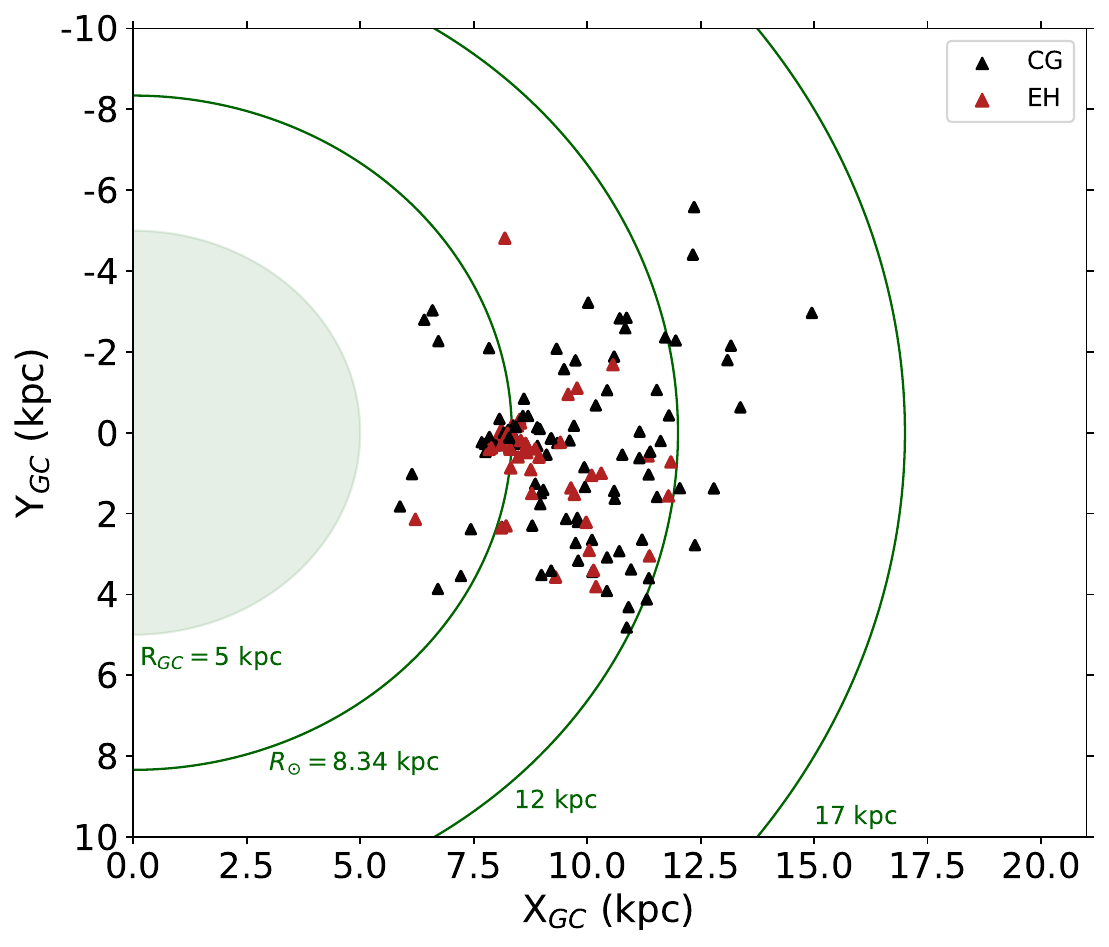}
 	\caption{ \small The OCCAM DR19 sample based on \citet{HUNT_2023} membership plotted in the Galactic plane. Black triangles are clusters that are present in the both the \citet{cg20} and \citet{HUNT_2023} based samples of open clusters, while red triangles show clusters that are in the \citet{HUNT_2023} based sample only.  The concentric circles show $R_{GC}$ = 5, 8.34 (the solar circle), 12, and 17 kpc.    }
 	\label{fig:XYfulleh}
 \end{figure}

In total using the EH proper motion member catalog resulted in $\sim800$ stars in 150 open clusters. In Figure \ref{fig:XYfulleh}, the open clusters are plotted in the Galactocentric X-Y plane, where you can see the clusters which are present in both of the \gaia-based proper motion catalogs (black triangles) and which show up only when using the EH catalog (red triangles). 
Overall, there are 98 clusters that are in common. Of the 164 clusters from the main body of the paper, 66 are lost when switching to the EH catalog. 
The primary reason why a cluster was lost when switching from using the \citet{cg20} catalog (CG catalog) membership to the EH catalog membership is not that the cluster does not appear in the EH catalog, but is because the membership for individual clusters is different in the CG catalog as compared to the EH catalog, as
stars that were included in the CG catalog are no longer considered members in the EH catalog. 
Additionally, of the 56 ``new'' clusters that we gain by using the EH catalog, 52 of them are single star clusters, with the majority being within the solar neighborhood. 
The reason for this is the open cluster open fiber program in SDSS-V/MWM, used to target new open cluster stars in SDSS-V, was targeted based on the CG catalog. 
For these reasons, we decided to stick with the CG catalog as the basis for the primary analysis. 
We present the overall metallicity gradients with just the EH catalog in Figure \ref{fig:eh_feh_grad} for completeness as both the CG and EH catalogs membership probabilities are reported in the VAC.

\subsection{Galactic Metallicity Gradients}

\begin{figure*}[t!]
 	\begin{center}
         \epsscale{1.05}
     \plotone{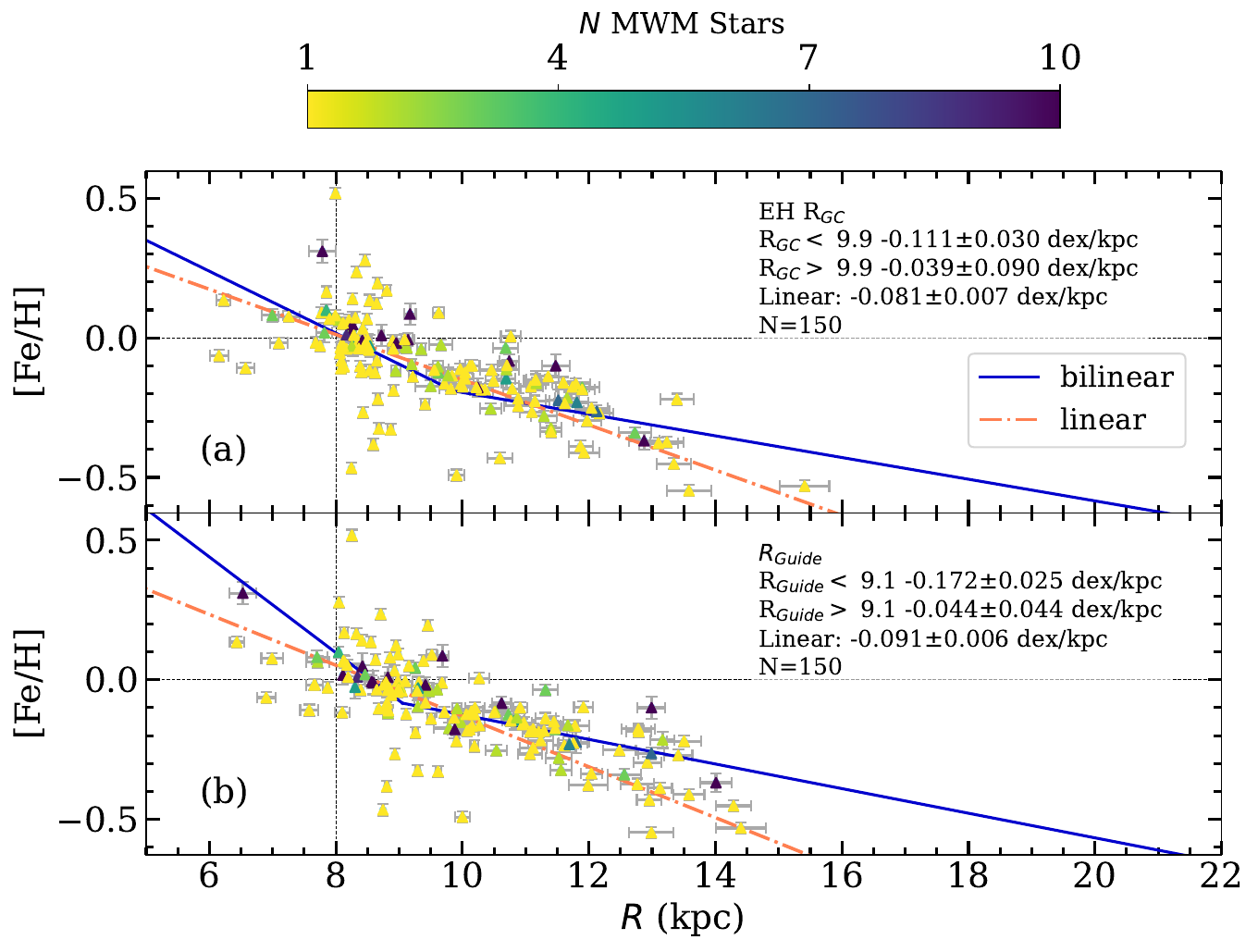} 
 	\end{center}
 	\caption{ \small The Galactic metallicity ([Fe/H]) gradients using the full sample of reliable clusters from the \citet{HUNT_2023} analysis (shown as triangles), as a function of current Galactocentric radius ($R_{GC}$; top panel (a)) and guiding center radius ($R_{Guide}$; bottom panel (b)). The bilinear fit (blue lines), as well as the linear fit (choral dot-dashed line) is shown. Fit parameters and knee locations are indicated within each panel. The color bar indicates the number of OCCAM member stars in each cluster, saturating at a value of 10 stars. }
 	\label{fig:eh_feh_grad}
 \end{figure*}

We compute both linear and bilinear fits to the EH catalog open cluster sample in radius vs [Fe/H] space. As in the main body of the paper, both Galactocentric radius (\rgc) and guiding center radius (\rguide) are used to compute the two fits, which are shown in Figure \ref{fig:eh_feh_grad}. With respect to $R_{GC}$, the linear gradient was calculated to be $-0.081 \pm 0.007 \text{ dex kpc}^{-1}$.
The inner slope of the bilinear fit was calculated to be $-0.111 \pm 0.030 \text{ dex kpc}^{-1}$, with the knee at $9.9 \pm 2.3$ kpc and an outer slope of $-0.039 \pm 0.090 \text{ dex kpc}^{-1}$. 
The slope of the linear fit with respect to \rguide was calculated to be $-0.091 \pm 0.006 \text{ dex kpc}^{-1}$. 
The bilinear fit has an inner slope of $-0.172 \pm 0.025 \text{ dex kpc}^{-1}$, with a knee at $9.1 \pm 1.3$ kpc and an outer slope of $-0.044 \pm 0.044 \text{ dex kpc}^{-1}$. 

\subsection{Comparison to the \citet{cg20} Cluster Sample Gradients}

In order to quantify how the metallicity ([Fe/H]) gradient changes when using the different membership catalogs we compare the gradients calculated in this appendix, using the EH catalog, with those in the main body of the text, using the CG catalog. 
When using \rgc, the gradients are consistent with each other. The CG catalog has a linear gradient of $-0.075 \pm 0.006 \text{ dex kpc}^{-1}$, which is within $1\sigma$ of the EH catalog linear gradient, $-0.081 \pm 0.007 \text{ dex kpc}^{-1}$. 
For the bilinear fit the inner slopes are very similar for the CG catalog and the EH catalog, at $-0.100 \pm 0.019 \text{ dex kpc}^{-1}$ and $-0.111 \pm 0.030 \text{ dex kpc}^{-1}$ respectively. The similarity continues for knee placements, $10.0 \pm 1.7$ kpc vs $9.9 \pm 2.3$ kpc, and outer slopes $-0.044 \pm 0.036 \text{ dex kpc}^{-1}$ vs $-0.039 \pm -0.090 \text{ dex kpc}^{-1}$.
Though we do note the increase in $1\sigma$ errors in all of these measurements.

The gradients with respect to $R_{Guide}$, however, are considerably different between the two catalogs. The linear slope using the CG catalog was calculated to be $-0.068 \pm 0.005 \text{ dex kpc}^{-1}$, but here with the EH catalog we calculated a steeper slope of $-0.091 \pm 0.007 \text{ dex kpc}^{-1}$. For the bilinear fit, the knee moves inwards almost 3 kpc, from $12.0 \pm 2.7$ kpc using the CG catalog to $9.1 \pm 1.3$, though due to the large uncertainties in these numbers, they are consistent with each other. 
The inner slope of the bilinear fit is where the largest change is seen. Using the CG catalog, a slope of $-0.072 \pm 0.020 \text{ dex kpc}^{-1}$ was calculated, but here we calculate an inner slope of $-0.172 \pm 0.025$, a 0.1 dex decrease. The outer slope using the CG catalog was determined to be $-0.015 \pm 0.085 \text{ dex kpc}^{-1}$ while the EH catalog slope was determined to be $-0.044 \pm 0.044 \text{ dex kpc}^{-1}$. While the values themselves are considerably different, it is worth noting that both are generally consistent with a flat trend line. It is unclear what is causing the \rguide slopes to differ significantly when using the different catalogs, while the \rgc slopes are in good agreement. 

\bibliography{Otto}{}
\bibliographystyle{aasjournalv7}



\end{document}